\documentclass[twocolumn]{aastex631}
\usepackage{graphicx}
\usepackage{import}
\usepackage{subfigmat}
\usepackage{paralist}
\usepackage{amsmath}

\defcitealias{Tanaka+18}{Paper I}

\shorttitle{The Missing Satellite Problem Outside of the Local Group. II.}
\shortauthors{Nashimoto et al.}

\begin{document}

\title{
The Missing Satellite Problem Outside of the Local Group. II. \\
Statistical Properties of Satellites of Milky Way-like Galaxies
}

\correspondingauthor{Masashi Nashimoto}
\email{nashimoto@astron.s.u-tokyo.ac.jp}

\author[0000-0002-1221-1708]{Masashi Nashimoto}
\affiliation{National Astronomical Observatory of Japan, Osawa 2-21-1, Mitaka, Tokyo 181-8588, Japan}

\author[0000-0002-5011-5178]{Masayuki Tanaka}
\affiliation{National Astronomical Observatory of Japan, Osawa 2-21-1, Mitaka, Tokyo 181-8588, Japan}
\affiliation{Department of Astronomical Science, The Graduate University for Advanced Studies, SOKENDAI, 2-21-1 Osawa, Mitaka, Tokyo, 181-8588, Japan}

\author[0000-0002-9053-860X]{Masashi Chiba}
\affiliation{Astronomical Institute, Tohoku University, 6-3, Aramaki, Aoba-ku, Sendai, Miyagi, 980-8578, Japan}

\author[0000-0002-8758-8139]{Kohei Hayashi}
\affiliation{National Institute of Technology, Ichinoseki College, Takanashi, Hagisho, Ichinoseki, Iwate, 021-8511, Japan}
\affiliation{Astronomical Institute, Tohoku University, 6-3, Aramaki, Aoba-ku, Sendai, Miyagi, 980-8578, Japan}
\affiliation{Institute for Cosmic Ray Research, The University of Tokyo, Chiba, 277-8582, Japan}

\author[0000-0002-3852-6329]{Yutaka Komiyama}
\affiliation{Dept. of Advanced Sciences, Faculty of Science and Engineering, Hosei University, 3-7-2 Kajino-cho, Koganei-shi, Tokyo 184-8584, Japan}

\author[0000-0003-0137-2490]{Takashi Okamoto}
\affiliation{Faculty of Science, Hokkaido University, N10 W8, Kitaku, Sapporo, Hokkaido 060-0810, Japan }

\begin{abstract}
We present a new observation of satellite galaxies around seven Milky Way (MW)-like galaxies located outside of the Local Group (LG) using Subaru/Hyper Suprime-Cam imaging data to statistically address the missing satellite problem.
We select satellite galaxy candidates using magnitude, surface brightness, S\'{e}rsic index, axial ratio, full width half maximum, and surface brightness fluctuation cuts, followed by visual screening of false-positives such as optical ghosts of bright stars.
We identify 51 secure dwarf satellite galaxies within the virial radius of nine host galaxies, two of which are drawn from the pilot observation presented in Paper I\@.
We find that the average luminosity function of the satellite galaxies is consistent with that of the MW satellites, although the luminosity function of each host galaxy varies significantly.  We observe an indication that more massive hosts tend to have a larger number of satellites.
Physical properties of the satellites such as the size-luminosity relation is also consistent with the MW satellites.
However, the spatial distribution is different; we find that the satellite galaxies outside of LG shows no sign of concentration or alignment, while that of the MW satellites is more concentrated around the host and exhibits a significant alignment.
As we focus on relatively massive satellites with $M_V<-10$, we do not expect that the observational incompleteness can be responsible here. 
This trend might represent a peculiarity of the MW satellites,
and further work is needed to understand its origin.
\end{abstract}


\section{Introduction} \label{sec:intro}  
The galaxy formation theory based on the Lambda cold dark matter (LCDM) model is a powerful cosmological model that can explain many astronomical observations. 
While the LCDM model reproduces the large-scale structure well, possible problems have been pointed out on a small scale; 
e.g., 
the cusp-core problem \citep{Moore94, Flores+94, McGaugh+01, Gilmore+07, KuziodeNaray+08},
the too-big-to-fail problem \citep{Boylan-Kolchin+11, Boylan-Kolchin+12, Parry+12},
the satellite alignment problem \citep{Ibata+13, Pawlowski+13, Pawlowski+15}, and
the angular momentum problem \citep{vandenBosch+01}.
In particular, the discrepancy in the number of dwarf galaxies in the Milky Way (MW) between theory and observation, known as the missing satellite problem, has remained unresolved for more than 20 years \citep{Kauffmann+93, Klypin+99, Moore+99}.

To resolve the missing satellite problem, high-resolution cosmological simulations accounting for realistic baryon physics have been performed, allowing for direct comparisons between observation and simulations \citep{Wetzel+16, Brooks+17, Fielder+19}.
\cite{Kim+18} argue that the MW no longer has the missing satellite problem once completeness corrections for faint dwarf galaxies are accounted for. 
Meanwhile, there is no consensus on whether the missing satellite problem is universal outside the MW or the Local Group (LG).
While model adjustments to reproduce the MW is fruitful, it may not fully resolve
the missing satellite problem (see, e.g.~\citealt{McConnachie12, Honma+19}) because the MW is not necessarily be a typical galaxy.
A survey of satellite galaxies outside the LG is crucial to statistically address the problem.

In recent years, the detection of dwarf satellite galaxies outside the LG has been reported one after another 
(e.g.,~\citealt{Chiboucas+09, Irwin+09, Stierwalt+09, Trentham+09, Kim+11, Ferrarese+12, Ferrarese+16, Ferrarese+20, Chiboucas+13, Sales+13, Crnojevic+14, Crnojevic+16, Crnojevic+19, Merritt+14, Spencer+14, Karachentsev+15, Muller+15, Muller+17, Muller+18, Muller+19, Muller+21, Munoz+15, Carlin+16, Carlin+21, Bennet+17, Bennet+19, Bennet+20,  Danieli+17, Geha+17, Park+17, Park+19, Smercina+17,  Cohen+18, Eigenthaler+18, Greco+18, Kondapally+18, Smercina+18, Venhola+18, Venhola+21, Carlsten+19, Carlsten+21, Carlsten+22, Zaritsky+19, Byun+20, Danieli+20, Habas+20, Muller+20, Davis+21, Drlica-Wagner+21, Garling+21, Mao+21, Prole+21, Su+21, Tanoglidis+21, LaMarca+22, Mutlu-Pakdil+22, Wu+22}).
\citet[hereafter \citetalias{Tanaka+18}]{Tanaka+18} investigated dwarf satellite galaxies around two nearby (15–20 Mpc) MW-like galaxies observed with the Subaru/Hyper Suprime-Cam (HSC). 
This paper is an extension of that work in two respects:
\begin{inparaenum}[(1)]
    \item we perform a more careful selection of dwarf satellite galaxies, and
    \item we address the above mentioned problems a larger sample of MW-like galaxies.
\end{inparaenum}
We demonstrate that a large statistical sample of satellite galaxies is essential to address them and the MW satellites are actually not typical in terms of their spatial distribution.

This paper is structured as follows.
We present the new observational data in Section \ref{sec:data}, followed by a development of the dwarf satellite identification scheme in Section \ref{sec:methods}.
We then examine physical properties of the satellite galaxies in Section \ref{sec:results}.
Finally, we discuss and summarize the results in Section \ref{sec:summary}.
We adopt the AB magnitude system \citep{oke83} throughout the paper. 
\section{Data} \label{sec:data}
The selection of host galaxies with MW-like mass is essentially the same as that adopted in \citetalias{Tanaka+18}.
We here give a brief review of it.
First, we use the 2MASS Large Galaxy Atlas \citep{Jarrett+03,Skrutskie+06} as the primary photometric data supllemented by the 2MASS extended source catalog \citep{Jarrett00} and distances from the HyperLEDA database \citep{Makarov+14} to infer stellar mass of nearby galaxies. 
When multiple distance measurements are available for a given galaxy, we adopt the average of them.  
The stellar mass is then translated into halo mass using the abundance matching from \citet{Moster+10}.  
Assuming the halo mass of the MW is $(1-2)\times10^{12}\,\rm{M_\odot}$ \citep{Bland-Hawthorn+16}, and also considering the uncertainty in the stellar mass and abundance matching, we select objects with halo mass of $(0.5-4)\times10^{12}\,\rm{M_\odot}$.

This is the primary constraint in our target selection scheme.  
We impose a further constraint that the virial radius, defined throughout this paper as the radius within which the mean interior density is 200 times the critical density, has to be about the size of the field of view of HSC.  
As a result of these constraints, our targets are typically located at $15-20$ Mpc. 
There are also other constraints applied to avoid bright stars and severe Galactic extinction (see \citetalias{Tanaka+18} for details).  
We do not explicitly apply any conditions on the presence of a massive neighbor (the MW has a massive companion), but we do exclude galaxie in obvious groups and clusters from the sample.

Our targets were observed over multiple semesters since 2018.  
We use the $g$ and $i2$ filters and expose 30 min in total in each filter.  
The dither length of 3 arcmin is applied between the individual exposures to fill the CCD gaps and reduce artifacts.  
Twenty objects were observed, but not all of them have a complete data set taken under acceptable conditions.
In this paper, we focus on seven objects observed in both filters under good conditions as summarized in Table \ref{tab:obj_prop}.  
Among them, N5866 actually does not satisfy one of the conditions above (i.e., its virial radius is larger than the field of view of HSC).  
It was a target of a prior run and was already partially observed, and we choose to include in the analysis here by accounting for the missing area of the virial radius.  In addition to the newly observed objects, we also use two objects observed in the pilot run \citepalias{Tanaka+18}.

The data were processed with \texttt{hscPipe} v8.4 \citep{Juric+17,Bosch+18,Bosch+19,Ivezic+19}.  
As described in \cite{Aihara+19}, this version is able to subtract the background sky on a large scale, which is ideally suited for the purpose of the paper as we focus on nearby galaxies.
The pipeline generates a multi-band photometric catalog, but we use the coadd images only and perform separate object detection and photometry optimized for extended dwarf galaxies as we detail below.

\begin{deluxetable*}{cccccccc}
\tablecaption{Properties of the targets. The last two objects are from the pilot run.}
\tablehead{
\colhead{Object} & \colhead{R.A.} & \colhead{Dec} & \colhead{Distance} & \colhead{Stellar Mass} & \colhead{Halo Mass} & \multicolumn{2}{c}{Virial Radius}\\
& \colhead{(deg)} & \colhead{(deg)} & \colhead{(Mpc)} & \colhead{($10^{10} M_\odot$)} & \colhead{($10^{11} M_\odot$)} & \colhead{(kpc)} & \colhead{(arcmin)}}
\startdata
N3338 & 160.53 & +13.75 & 23.38\tablenotemark{a}   & 5.77 & 22.9 & 266.4 & 39.2 \\
N3437 & 163.15 & +22.93 & 22.93\tablenotemark{a,b} & 4.70 & 17.4 & 243.0 & 36.4 \\
N5301 & 206.60 & +46.10 & 20.58\tablenotemark{a,b} & 1.77 & 6.76 & 177.4 & 29.6 \\
N5470 & 211.60 & +06.03 & 20.32\tablenotemark{a}   & 1.41 & 5.62 & 166.8 & 28.2 \\
N5690 & 219.43 & +02.28 & 19.07\tablenotemark{a,c} & 1.98 & 7.24 & 181.5 & 32.7 \\
N5866 & 226.63 & +55.77 & 14.79\tablenotemark{b,d} & 7.13 & 33.1 & 301.2 & 70.0 \\
N7332 & 339.35 & +23.80 & 23.01\tablenotemark{c}   & 3.23 & 11.2 & 210.0 & 31.4 \\
\hline
N2950 & 145.65 & +58.90 & 14.93\tablenotemark{c}   & 1.72 & 6.61 & 176.0 & 40.5 \\
N3245 & 156.83 & +28.50 & 20.89\tablenotemark{c}   & 4.04 & 14.1 & 226.7 & 37.3 \\
\enddata
\tablerefs{\textsuperscript{a}\cite{Sorce+14}; \textsuperscript{b}\cite{Tully+08}; \textsuperscript{c}\cite{Tonry+01}; \textsuperscript{d}\cite{Ciardullo02+}.}
\label{tab:obj_prop}
\end{deluxetable*}
\section{Methods} \label{sec:methods}
This section describes the identification scheme of dwarf satellite galaxies.
In short, we make the following steps:
\begin{enumerate}
   \setlength{\itemsep}{-1mm}
   \setlength{\leftskip}{-1mm}
   \item Object detection with Source Extractor to make a rough selection of extended dwarf satellites.
   \item Detailed measurements with GALFIT to carefully screen the candidates.
   \item Surface brightness fluctuation distance measurements to eliminate object located far from the host galaxies.
   \item Visual inspection to exclude any remaining artifacts and grade the candidates.
\end{enumerate}
We detail each of these steps in what follows.

\subsection{Object Detection with Source Extractor}
\label{sec:SExtractor}
To begin with, we detect objects on the coadd images with Source Extractor \citep{SExtractor96}.
We use the $i$-band images for detection because they have slightly better seeing than the $g$-band. 
There is no essential difference in the detection results between the $i$-band images and the $g$-band ones.
The detection threshold is set to $\mathrm{DETECT \_ THRES}=1\sigma$, and the objects with more than 50 pixels exceeding the threshold are detected ($\mathrm{DETECT \_ MINAREA}=50$ pix).
These parameters are similar to those adopted in \citetalias{Tanaka+18} but with a more conservative number of connected pixels so that we do not miss relatively compact dwarf galaxies.
The size of the detected objects corresponds to a diameter of $\sim150$ pc at $d=23$ Mpc, which is the largest distance in our sample.  This is sufficiently small to detect all satellite galaxies with $M_V \lesssim -10$ as shown in Figure 3 of \citetalias{Tanaka+18}, which used a more exclusive detection parameter.

In order to reduce false detection due to contamination caused by bright stars, we create masks around bright stars and exclude objects in the masked regions. 
We generate the mask around bright stars in the same way as \citetalias{Tanaka+18}. Bright stars have many saturated pixels at the center.  
We thus first search for groups of saturated pixels in the coadd images, identify their approximate centers and define circular regions around them whose sizes correlate with the number of saturated pixels.  
The size is chosen empirically to make a sufficiently large region around the star.  
A rectangular mask is also defined to exclude a region affected by the bleeding trail.
Since the virial radius of N5866 is larger than the field of view of the HSC (see Table \ref{tab:obj_prop}), all regions outside the field of view are treated as masked when comparing it with other galaxies.
In addition, there is a group of galaxies in the northeastern part of N3338.
Spectroscpic redshifts from the Sloan Digital Sky Survey \citep{Ahumada+20} indicates that this is a background galaxy group (see Figure \ref{fig:images}).  
We apply an additional mask to exclude this group.

\begin{figure}[t]
    \centering
	\includegraphics[scale=1]{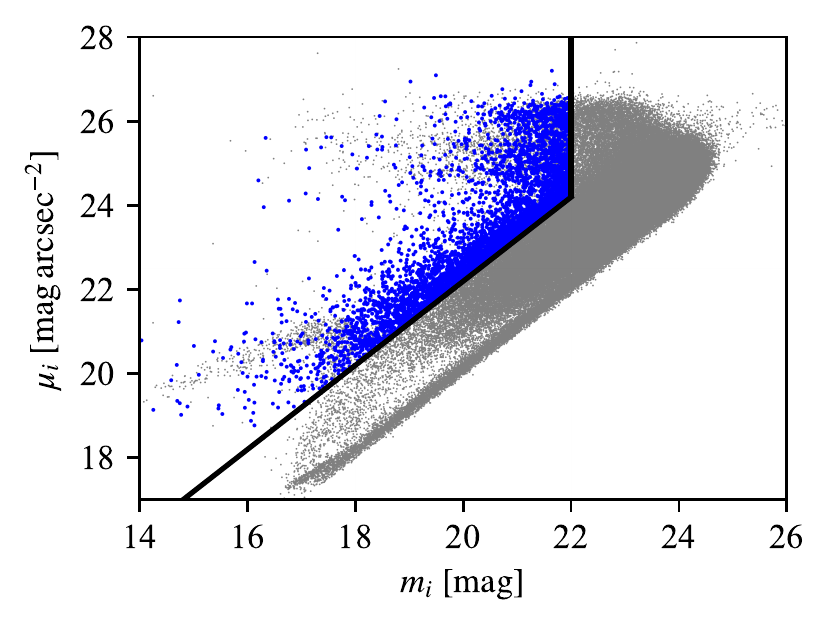}
	\caption{
	    Surface brightness versus magnitude for all detected sources around N5866. 
	    The solid lines represent the surface brightness and magnitude cuts. 
	    The blue (gray) dots represent objects that satisfy (do not satisfy) the surface brightness, magnitude, and axial ratio criteria and are outside the masked regions.
	    }
    \label{fig:sextractor}
\end{figure}
We then apply surface brightness and magnitude cuts to select candidates of dwarf satellite galaxies.  
To be specific, we apply a surface brightness cut as $\mu_i = m_i + 2.5\log(2\pi r_\mathrm{eff}^2)$ where $r_\mathrm{eff}=1.55$ arcsec, which roughly corresponds to 150 pc at $d=23$ Mpc.
\citetalias{Tanaka+18} applied a more stringent cut but they still achieved a high completeness.
As our cut here is more inclusive, we can expect a higher completeness.
In addition, we apply a magnitude cut of $m_i \leq 22$ mag, which is roughly equivalent to $M_V \lesssim -10$ mag.
The masks described above cannot completely remove false detection, but we find that the fake sources often have very high ellipticities.  
We thus impose an axial ratio cut $\epsilon \geq 0.1$ to remove them at this point.
Figure \ref{fig:sextractor} visualizes these surface brightness and magnitude cuts.
The two sequences around $\mu_i=20$ and 26 $\mathrm{mag\, arcsec^{-2}}$ in the upper left of this figure are dominated by false detection.
We show N5833 as an example here but the other host galaxies are similar.
The total number of detected objects among seven host galaxies is 1,385,854, and 45,365 objects satisfy the cuts.

\subsection{Measurements by GALFIT}
\label{sec:galfit}
The dwarf satellite candidates we have at this point include a lot of contaminant objects such as optical ghosts of the camera.  
In order to further eliminate them, we measure fluxes and shapes of the sources more precisely using GALFIT \citep{galfit02, galfit10}.
For this, we follow a multi-step procedure.  
First, we generate an image cutout for each candidate selected from the surface brightness versus magnitude cut (Figure \ref{fig:sextractor}).  
We then run Source Extractor to detect sources and measure their centers, position angles, and approximate sizes.
Note that the measurement is performed not only for the candidates, but for sources around them as well.  
The detection parameters are tuned to detect all sources;
$\texttt{DETECT\_THRES}=1.5\sigma$ and $\texttt{DETECT\_MINAREA}=8$ pix.
GALFIT is finally run with these measurements as a initial guess.  
GALFIT successfully fits the sources in many cases, but there are cases where manual interventions are needed such as a dwarf galaxy candidate located close to a bright star.  
Manual masking of neighboring sources are necessary in many of these cases.
GALFIT failures on objects that are clearly not dwarf galaxies from an visual inspection are simply ignored from the following analyses.

\begin{figure*}[t]
    \centering
	\includegraphics[scale=1]{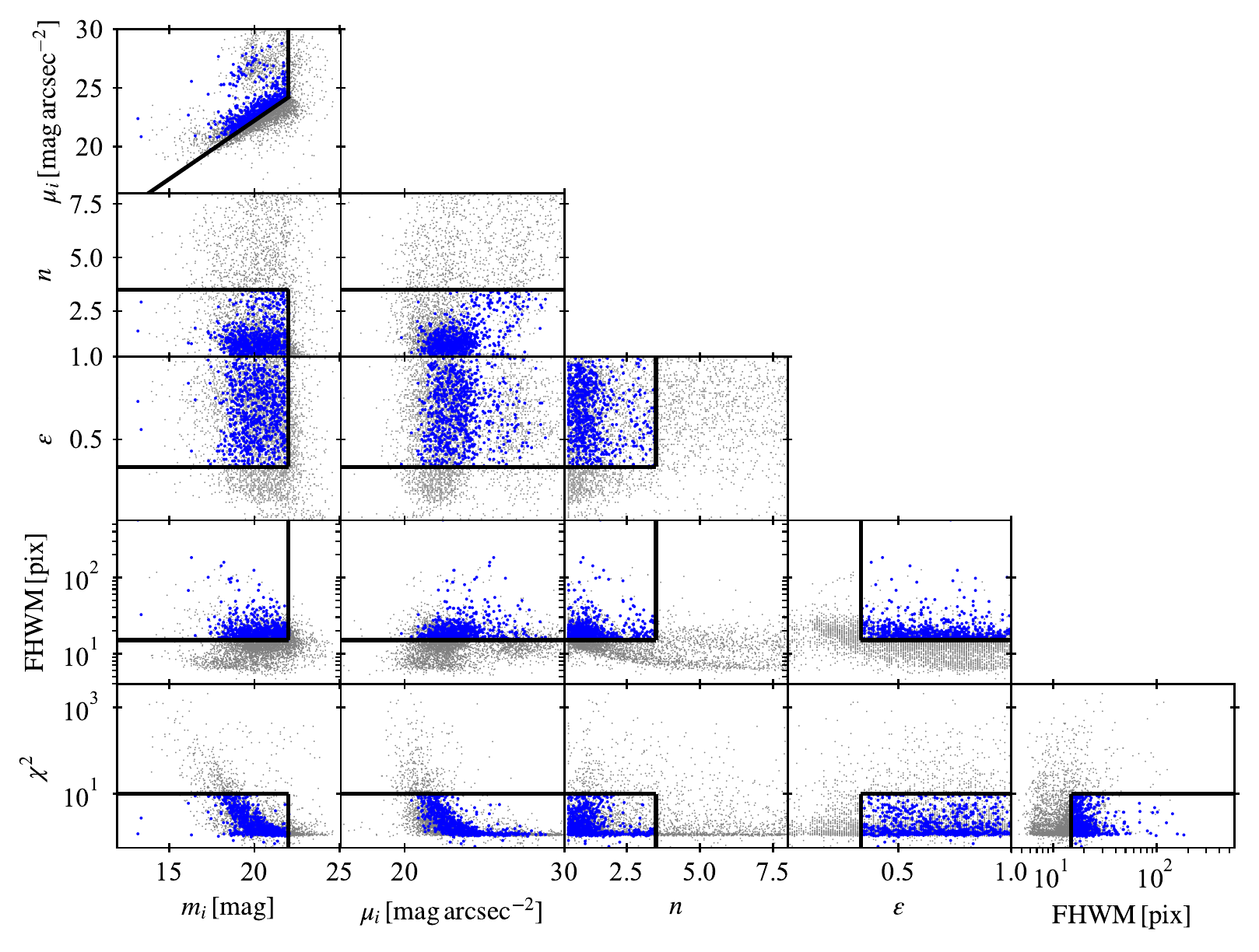}
	\caption{
    Measured properties of the candidates around N5866.
    The dots are the candidates after the initial Source Extractor cuts (Section \ref{sec:SExtractor}).
	The solid lines represent each parameter cut.
	The blue dots are the objects that satisfy all the criteria.
	}
    \label{fig:galfit}
\end{figure*}
\twocolumngrid
We filter the candidates by imposing cuts on six parameters from GALFIT measurements;
magnitude ($m_i$), surface brightness ($\mu_i$), S\'{e}rsic index ($n$), axial ratio  ($\epsilon$), full width at half maximum (FWHM), and reduced chi-squares measured in the central $25\times25$ pixels of the residual image ($\chi^2$).
The magnitude and surface brightness cuts are the same as those in the first selection (see Section \ref{sec:SExtractor}), but we now use GALFIT measurements.
For the other parameters, we apply $n\leq3.5$, $\epsilon\geq1/3$, $\mathrm{FWHM}\geq15\,\mathrm{pix}$, and $\chi^2\leq10$, to down-select the candidates.
These thresholds are empirically set; we first visually identify dwarf satellite candidates around N5866 and then choose the thresholds to include all of them, while excluding obvious outliers.  
In other words, these cuts exclude only obvious non-dwarf galaxies rather than to narrow-down the candidates.
Figure \ref{fig:galfit} summarizes the cuts.
As seen in the upper left panel, there are objects that do not satisfy the surface brightness and magnitude cutoffs in the GALFIT measurements, although they do with the Source Extractor measurements.  
This is because of improved source detection and deblending in GALFIT\@. 
Our source detection with Source Extractor is tuned to identify extended sources and we are missing compact sources, which may contaminate outskirts of nearby objects.  
In the GALFIT run, we detect essentially all sources in the image cutouts and fit them all, which allows us to deblend sources properly.
In the case of N5866, 678 objects satisfy the GALFIT cuts. 
The same criteria are applied to all the other galaxies, leaving 4,917 dwarf galaxy candidates in total.

We note that the cuts are optimized for dwarf satellites; as seen from the lower left panel of Figure \ref{fig:galfit}, some bright galaxies are not well fitted by the model and do not satisfy the chi-square cut due to significant structure. 
Fortunately, many of these galaxies have measured spectroscopic redshifts, and we prioritize the spectroscopic redshift results in our sample selection. 
That is, if the spectroscopic redshift is consistent with being a satellite, the galaxy is included in the satellite galaxy sample.
We may still miss bright satellite galaxies without spectroscopic data, but the do not affect our conclusions because the number of faint satellites is much larger and they dominate our statistical analyses, as seen in later discussions.

\subsection{Surface Brightness Fluctuation Distance}
\label{sec:sbfd}
The dwarf satellite candidate catalog now has much less contaminants but it is not completely clean, e.g., there is still contamination of foreground and background galaxies.
In order to remove them, we make an attempt to measure their distances using the surface brightness fluctuation (SBF) proposed by \cite{Tonry+88}.

Based on the method by \cite{Carlsten+19}, we measure the SBF distance in the following manner.
First, we subtract the best-fit GALFIT models of a satellite candidate and objects around it from the cutout image.
Then, we divide the residual image by the square root of the best fit model.
Next, we mask and exclude the region outside of $r_\mathrm{eff}$ of the candidate. 
In addition, to avoid contamination, we also exclude the area where neighboring objects overlap.
The masked image is then Fourier transformed and azimuthally averaged in the wavenumber space to obtain a one-dimensional power spectrum.
The same procedure is performed on the point spread function (PSF) image to calculate the PSF power spectrum, which is further convoluted with that of the masked image.
We estimate the SBF flux at $k=0$ and the photon shot noise, $P_0$ and $P_1$, by linearly fitting the SBF power spectrum with the convoluted PSF power spectrum using the following equation,
\begin{eqnarray}
    P_\mathrm{SBF}(k) 
    = P_0 P_\mathrm{PSF}(k) + P_1.
    \label{eq:ps_sbf}
\end{eqnarray}
The apparent magnitude of SBF ($m_i^\mathrm{SBF}$) is obtained by substituting into the following conversion formula,
\begin{eqnarray}
    m_i^\mathrm{SBF}
    = -\frac{5}{2} \log \left( \frac{P_0}{t_\mathrm{exp}} \right)
      + \mathrm{z.p.}, 
    \label{eq:mag_sbf}
\end{eqnarray}
where $t_\mathrm{exp}$ is the exposure time set at $t_\mathrm{exp}=1$\,sec, and the zero point is $\mathrm{z.p.} = 27$ in our case.
The absolute magnitude of SBF ($M_i^\mathrm{SBF}$) is estimated using the linear relationship between the SBF magnitude and color provided by \cite{Carlsten+19},
\begin{eqnarray}
    M_i^\mathrm{SBF}
    = (-3.17 \pm 0.19) + (2.15 \pm 0.35) \times (g-i).
    \label{eq:Mag_sbf}
\end{eqnarray}
We calculate the SBF distance ($d_\mathrm{SBF}$) using the definition of absolute magnitude,
\begin{eqnarray}
    m_i^\mathrm{SBF} - M_i^\mathrm{SBF}
    = \frac{5}{2} \log 
    \left( \frac{d_\mathrm{SBF}}{10\,\mathrm{pc}} \right)^2.
    \label{eq:d_sbf}
\end{eqnarray}

\begin{figure}[t]
    \centering
	\includegraphics[scale=1]{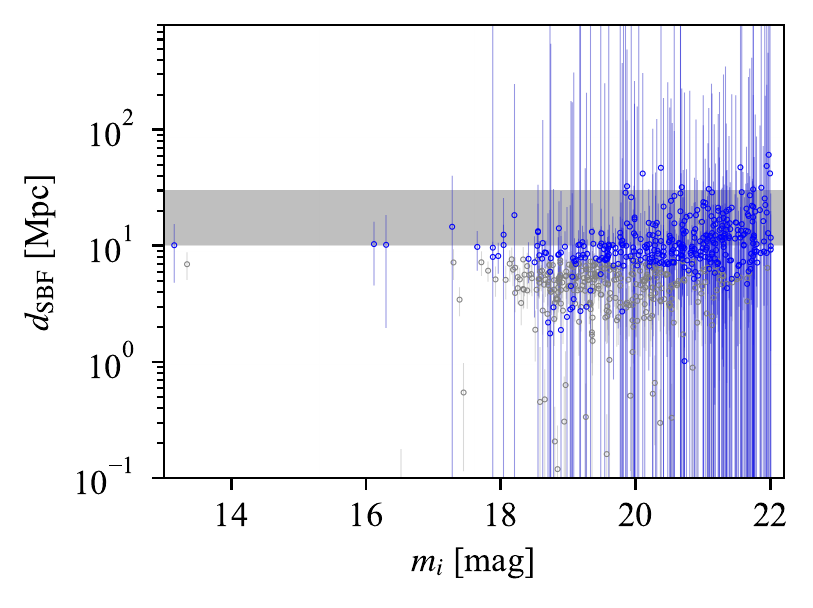}
	\caption{
	SBF distance versus magnitude for candidate objects around N5866. 
	The shaded region is the extent of the SBF distance cut.
	Blue (gray) circles represent the objects satisfying (not satisfying) the SBF distance criteria.
	}
    \label{fig:sbfd}
\end{figure}
Since the target galaxies are at a distance of about 20 Mpc (see Table \ref{tab:obj_prop}), objects whose SBF distance overlaps with 10-30 Mpc within the 1$\sigma$ error range are considered as candidates, taking into account the uncertainty of the SBF distance.
Figure \ref{fig:sbfd} shows the cut for N5866.
For N5866, 366 objects satisfy the SBF distance cut. 
For other galaxies, the number of candidates is reduced to about half, to a total of 2,737.

As can be seen in Figure \ref{fig:sbfd}, there are many objects located closer than 10 Mpc.
Some of these objects are due to prominent structure such as spiral arms and star forming regions, which enhances the SBF signal and thus the distance is underestimated.
But, some may be due to inaccurate GALFIT model; if the S\'{e}rsic model does not describe the galaxy well, we then underestimate its distance due to residual structure.  
As it is not trivial to evaluate the validity of the GALFIT model, we instead make an empirical approach to estimate how the possible underestimated distances affect the selection of the dwarf satellites.
We first visually check all the dwarf satellite candidates around a host galaxy, N5866, exclude obvious contaminants, and then perform more careful GALFIT analyses for 34 candidates and estimate their SBF distances.
We find that the final candidates are nearly identical; a few objects differ between the two catalogs, but they are all classified as possible satellites, not secure ones (see their definitions below).  
As such, they do not significantly affect the main conclusions of this work. 
The advantage of running the SBF distance cut first is that it significantly reduces the number of objects to visually check.  
As the visual classification is always subjective at least to some extent, we prefer to minimize it by applying the SBF distance cut.

\subsection{Visual Inspection}
Finally, the image of every candidate is visually checked to remove obvious contamination and also to give a confidence flag.
We classify the candidates into two classes: secure and possible satellite, depending on the confidence level of the classification.
To reduce the human bias, the two authors (MN and MT) perform the visual inspection independently, and the results of their classifications are combined into the final classification. 
Objects classified as secure by both are classified as secure dwarf satellites.
When one author judges an object as a secure candidate and the other classifies it as a possible one, we keep such objects as possible dwarf satellites.
An object that either one determines to be a contaminant is classified as a contaminant, even if the other judges it to be a secure candidate.
The number of the dwarf satellites for each host is shown in Table \ref{tab:catalog}. 
\begin{deluxetable}{ccccc}
\tablecaption{Number of secure and possible satellites.}
\tablehead{\colhead{Host Galaxy} & \multicolumn{2}{c}{Secure Satellite} & \multicolumn{2}{c}{Possible Satellite} \\
& $\leq r_{200}$ & $> r_{200}$ & $\leq r_{200}$ & $> r_{200}$}
\startdata
N3338 &  8 &  1 &  5 &  2 \\
N3437 &  5 &  1 &  4 &  6 \\
N5301 &  1 &  1 &  4 & 16 \\
N5470 &  0 &  1 &  5 & 10 \\
N5690 &  3 & 12 &  5 & 12 \\
N5866 &  9 &  0 & 10 &  0 \\
N7332 &  3 &  1 &  3 &  8 \\
\hline
N2950 &  9 &  0 &  4 &  0 \\
N3245 & 13 &  0 &  2 &  0 \\
\hline
Total & 51 & 17 & 42 & 54 \\
\enddata
\tablecomments{The columns denote satellite candidates with the projected distance from the central galaxy inside and outside of the virial radius, $r_{200}$, {shown in Table \ref{tab:obj_prop}}.}
\label{tab:catalog}
\end{deluxetable}

\subsection{Completeness Correction} \label{sec:completeness}
While the dwarf satellite catalog constructed here is now clean and robust, it suffers from incompleteness effects.
\citetalias{Tanaka+18} estimated the detection completeness of dwarf galaxies by inserting fake objects with a wide range of magnitudes and sizes in coadd images and repeating the detection.
It turned out that our detection method is indeed sensitive to diffuse galaxies.  If we put the local group satellites at the distance of our host galaxies,
we can detect essentially all satellites down to $M_V\sim10$ (except for M32, which is extremely compact).
Furthermore, the detection rate is approximately constant regardless of magnitudes and sizes of satellites.
Motivated by this, we perform a two-step completeness correction in this work using results from \citetalias{Tanaka+18} as our data in this paper are similar to those used in \citetalias{Tanaka+18} in terms of depth and seeing.

First is to correct for masked area; \citetalias{Tanaka+18} found that roughly 90\% of the incompleteness is due to the mask around bright stars; dwarf galaxies inside the mask simply cannot be detected.
As we have generated the mask for each host galaxy separately, the amount of correction is different for different host.   
The correction is done in a statistical manner; if 10\% of the area inside the virial radius is masked, then each satellite candidate will have a statistical weight of $1/0.9=1.11$.
There is a small level of remaining incompleteness ($\sim10\%$) after the masked area is accounted for.
It is likely due to object blending.  When a dwarf satellite is close to or overlapping with a nearby object, we may miss it.
This effect seems to be independent of host galaxies, which indicates that the blending with background galaxies is likely the primary cause.
We thus apply a 10\% correction factor for satellites around all hosts to account for this effect.
Whether we apply this additional correction or not does not alter the conclusions of the paper.

To summarize, our completeness correction is two-fold; we first account for the masked area and then correct for the residual 10\% incompleteness due to object blending.  All this is done in a statistical manner.
When considering the spatial distribution of satellite galaxies in Section \ref{sec:spa_dist}, the mask is accounted for as a function of the projected distance or angle.

\section{Results} \label{sec:results}
Based on the satellite galaxy sample we have built, we now present properties of the dwarf satellites within the virial radius of each of the nine MW-like host galaxies and compare them with observations inside and outside of the LG and also with the prediction from numerical simulations.
We note that two of the nine host galaxies are drawn from the pilot observation \citepalias{Tanaka+18}, and three dwarf satellites from the pilot data do not satisfy our dwarf galaxy selection criteria shown in Figures \ref{fig:galfit} and \ref{fig:sbfd}.
Two of them are rather bright galaxies and our selection criteria for dwarf galaxies are not optimal for them.
They both have spectroscopic redshifts from the Sloan Digital Sky Survey \citep{Ahumada+20} that are consistent with being satellite galaxies and we include them in the following analyses.
The other one is classified as a possible dwarf galaxy and we exclude it to be consistent with the current selection scheme.
Note that there are dwarf galaxies located outside of the virial radius of the host, but we focus only on those inside the virial radius.
The satellite galaxy catalog and images we use in the following analyses are given in Appendix \ref{sec:catalog}.

\subsection{Luminosity Function}
\label{sec:LF}
We start with the luminosity function of our dwarf galaxy sample. 
To compare with other studies, the absolute magnitude of $i$-band is converted to $V$-band according to the following formula, 
\begin{eqnarray}
    M_V = M_i - 0.036 - 0.592 \times (g-i).
    \label{eq:Mv}
\end{eqnarray}
This is based on the stellar population synthesis model of \cite{Bruzual+03}; we assume galaxies with exponentially declining star formation histories with various decay timescales formed at $z_f=10$ with subsolar metallicity ($Z=0.004$).  
We fit a linear function to $V-i$ as a function of $g-i$.

\begin{figure}[t]
    \centering
	\includegraphics[scale=1]{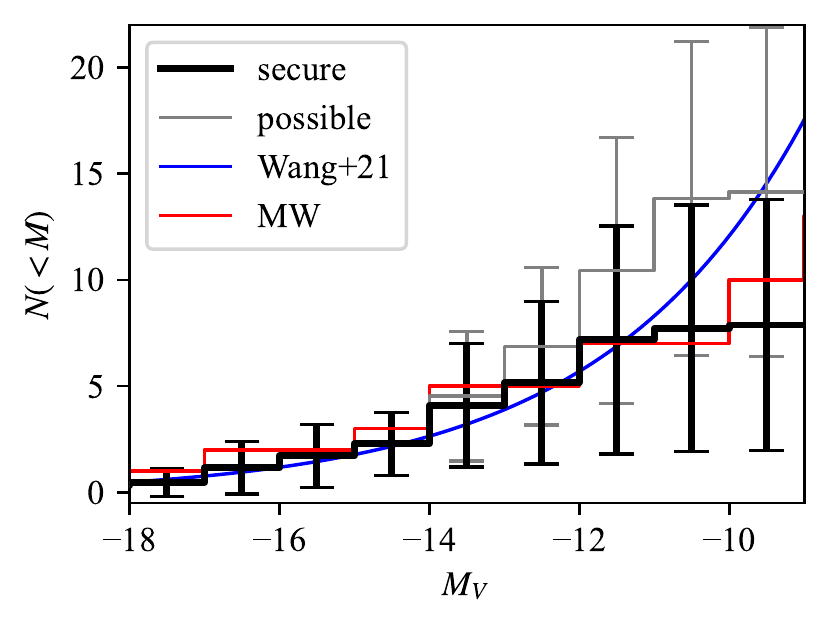}
	\caption{
	    Cumulative luminosity function of the detected dwarf galaxies and MW satellites \citep[red line;][]{McConnachie12}.
	    The black line includes only the secure satellites, while the gray line includes both secure and possible satellites.
	    The error bars represent the standard deviation among the nine host galaxies.
	    Note that the completeness correction (see Section \ref{sec:completeness}) is applied here.
	    The blue curve is the double Schechter function provided by \cite{Wang+21}, which is estimated by higher redshift dwarf galaxies.}
    \label{fig:LF}
\end{figure}
Figure \ref{fig:LF} shows the sample average of the cumulative luminosity function of dwarf satellites with projected distances within the virial radius.
The standard deviation represented by the error bars is comparable to the mean value, which indicates that the luminosity function has a wide range of variations for each MW-like galaxy.
The luminosity function of the secure satellites is comparable to that of the MW satellite galaxies.
If we include possible satellites, there are about twice as many satellites at the faint end. However, the possible satellites may include contamination and their luminosity function should be considered an upper bound.

For comparison with other galaxies, we overlay the double Schecter function,
\begin{eqnarray}
    \Phi(M) dM &=&
    \frac{2}{5} \ln 10 \sum_{i=1}^2 \Phi_{*,i} 10^{-0.4(M-M_0)(\alpha_i+1)}
    \nonumber \\
    && \times
    \exp \left( -10^{-0.4(M-M_0)} \right) dM,
    \label{eq:schecter}
\end{eqnarray}
where the values of the parameters, $M_0$, $\Phi_{*,i}$, and $\alpha_i$, are provided by \cite{Wang+21}, which are estimated based on dwarf galaxies around MW-like galaxies at higher redshifts from $z=0.001$ to $z\sim0.4$. 
The curve deviates from the luminosity function of the secure dwarf galaxies above $M_V \sim -11$, but it is within the scatter.
It is more consistent with our luminosity function including the possible satellites.

\begin{figure}[t]
    \centering
	\includegraphics[scale=.95]{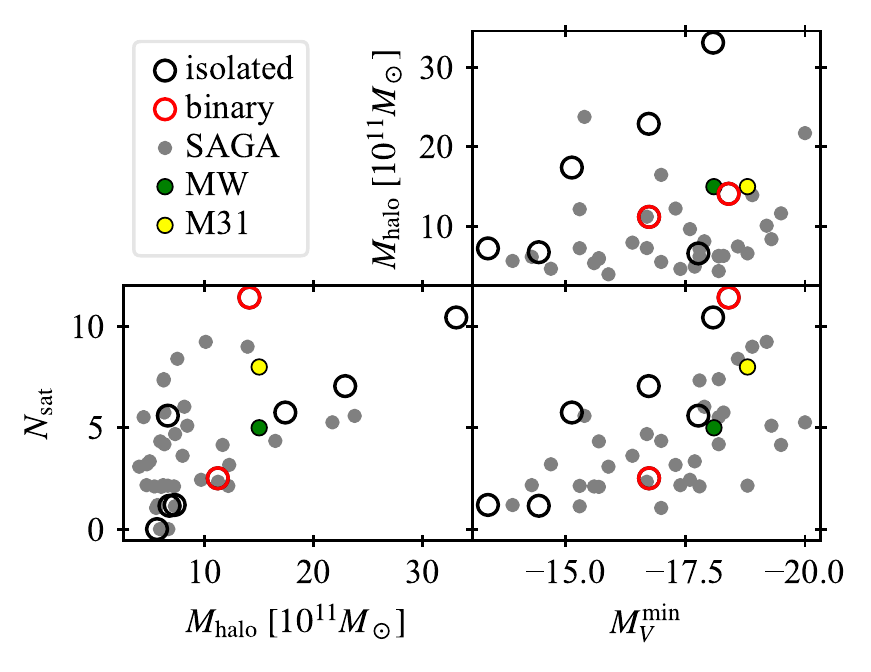}
	\caption{
	    The number of detected satellites versus host halo mass and magnitude of the brightest satellite.
	    The open circles are the secure satellites with $M_V < -12.1$  around the binary (red: N7332 and N3245) and isolated (black: others) host galaxy.
	    In the right panels, N5470 is excluded because N5470 has no secure satellite. 
	    The gray dots show the satellite galaxies outside the LG detected by the SAGA survey \citep{Mao+21}.
	    The number of satellite galaxies is corrected for completeness. 
	    The green and yellow dots represent the MW and M31, respectively.  We assume both galaxies have a halo mass of $\sim 1.5\times10^{12}\,M_\odot$ \citep{McConnachie12}.
	    }
    \label{fig:Nsat_Mhalo}
\end{figure}
One striking trend in Figure \ref{fig:LF} is the large host-host scatter in the observed luminosity function.
This scatter might be caused by the dependence on the host halo mass; we have targeted host galaxies with halo mass between 0.5 and $4\times10^{12}\ M_\odot$. 
To investigate the halo mass dependence, we show in Figure \ref{fig:Nsat_Mhalo} shows the relation between the number of detected secure satellites brighter than $M_V=-12.1$ and the host halo mass.
The magnitude cut here is applied to compare with a literature result (see below).
The figure indicates that the number of satellites increases as the host halo mass becomes larger.
The Pearson product-moment correlation coefficient between the number of satellites and the host halo mass is actually 0.73. 
For comparison, the satellite galaxies with $M_r<-12.3$ outside the LG detected by the SAGA survey are overlaid \citep{Mao+21}. 
Here we assume $M_V-M_r \simeq 0.2$, which is obtained in the same manner as in Equation \eqref{eq:Mv} using a typical color value.
The number of satellites is corrected by the spectroscopic coverage with the SAGA survey, and the host halo masses are given by the 2MRS group catalog \citep{Lim+17}.
The correlation coefficient of the SAGA result is 0.30, and when both samples are combined, the correlation coefficient is 0.51, making the correlation slightly weaker.
The figure seems to indicate that the observed scatter is due to combination of the halo mass dependence and host-host scatter at fixed mass.

\cite{Smercina+22} found the relationship between the number of satellites and the mass of the most massive satellite galaxy. This trend is similar to the relation between the number of satellites and the host halo mass, shown in the lower left panel, because the host halo mass is likely correlated to the mass of the most massive satellite. The lower right panel shows the relation between the number of satellites and the magnitude of the brightest satellite, which corresponds to the mass of the most massive satellite. As expected, this figure indicates the similar tendency.
These trends in the lower panels are consistent with the MW and M31.  
The number of satellites is also related to its environment \citep{Bennet+19}. N3245, whose halo mass is $M_\mathrm{halo}=14.1\times10^{11}\, M_\odot$, has a relatively large number of satellite galaxies, which may be because N3245 is a binary galaxy.
In any case, it is fair to conclude that the MW luminosity function is within the scatter of our luminosity function.

\begin{figure*}[t]
    \centering
    \begin{subfigmatrix}{4}
    \subfigure[N3338]{\includegraphics[scale=0.2]{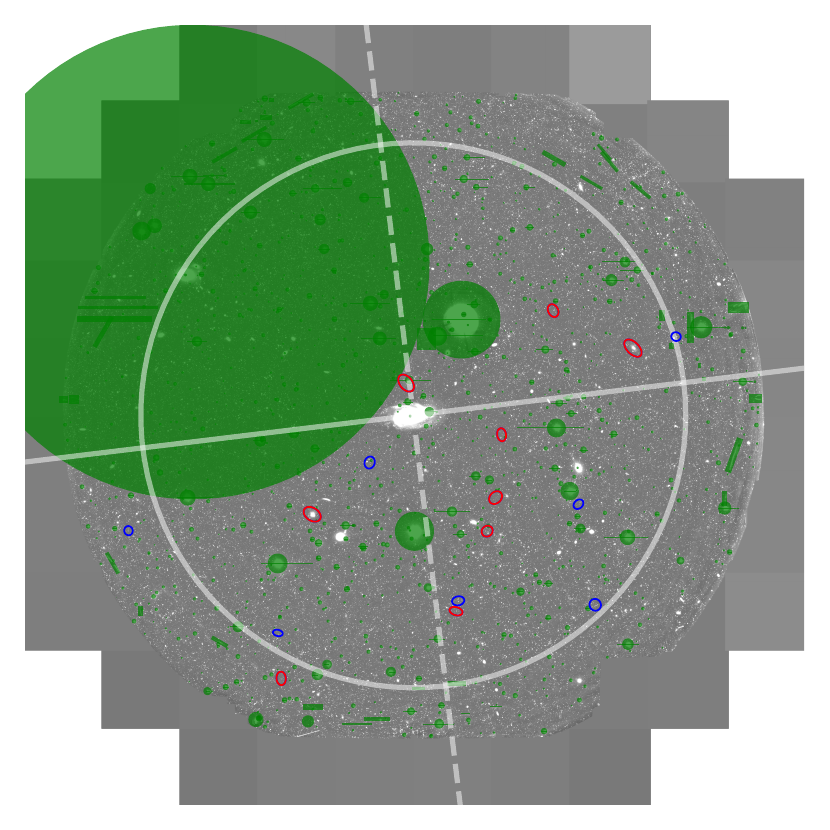}}
    \subfigure[N3437]{\includegraphics[scale=0.2]{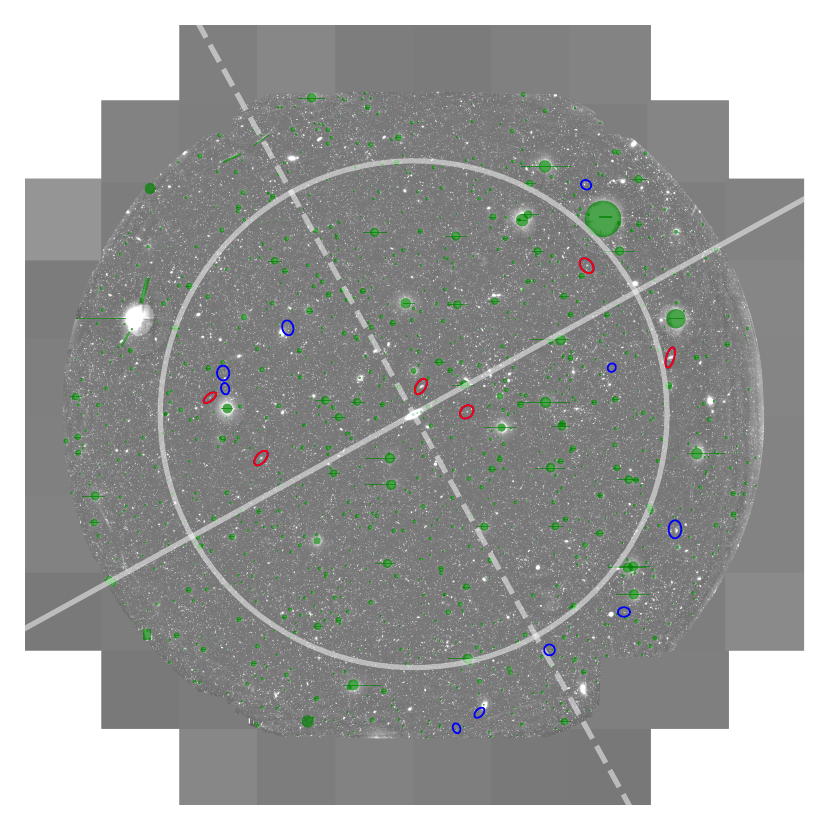}}
    \subfigure[N5301]{\includegraphics[scale=0.2]{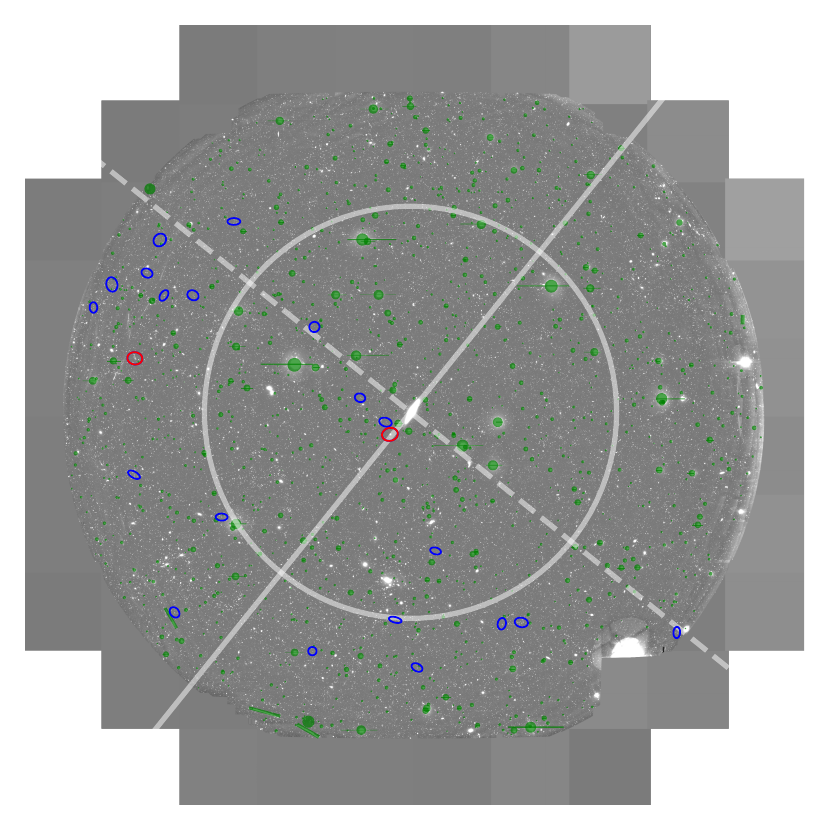}}
    \subfigure[N5470]{\includegraphics[scale=0.2]{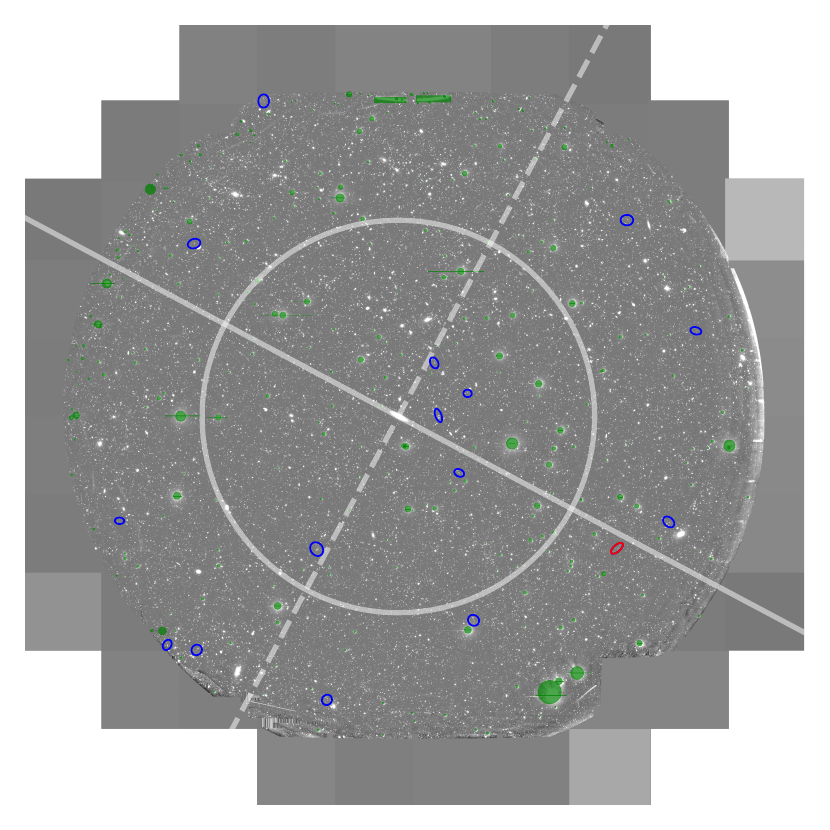}}
    \subfigure[N5690]{\includegraphics[scale=0.2]{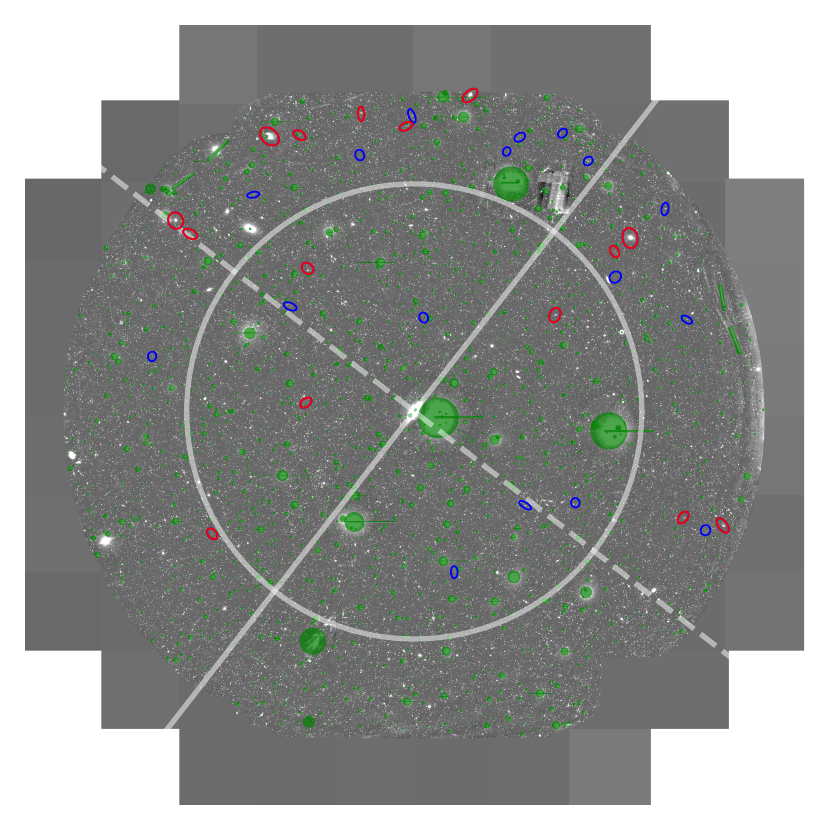}}
    \subfigure[N5866]{\includegraphics[scale=0.2]{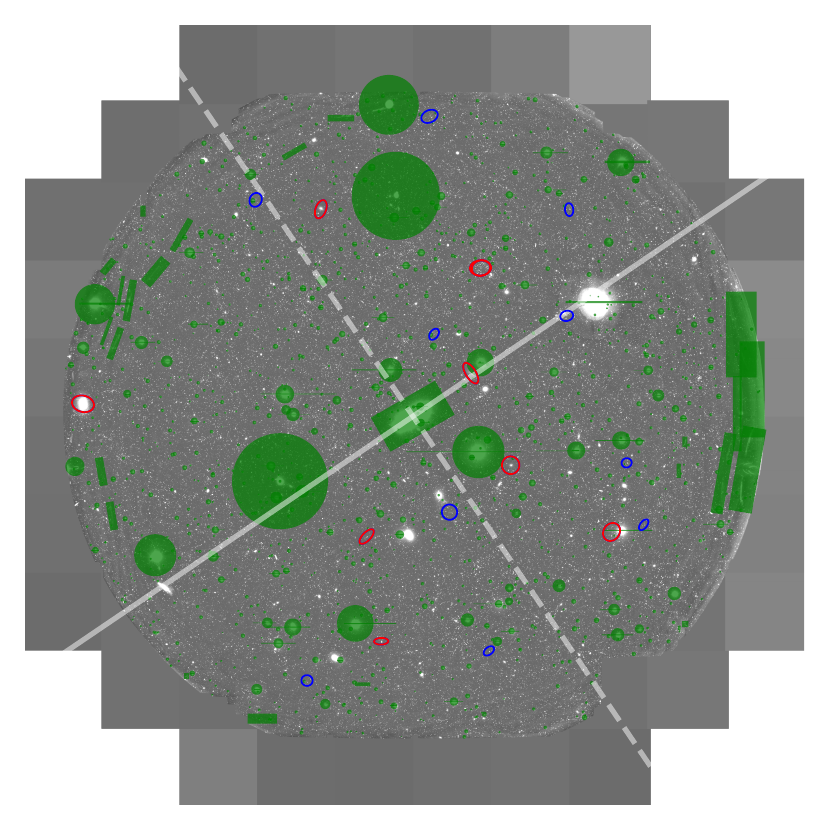}}
    \subfigure[N7332]{\includegraphics[scale=0.2]{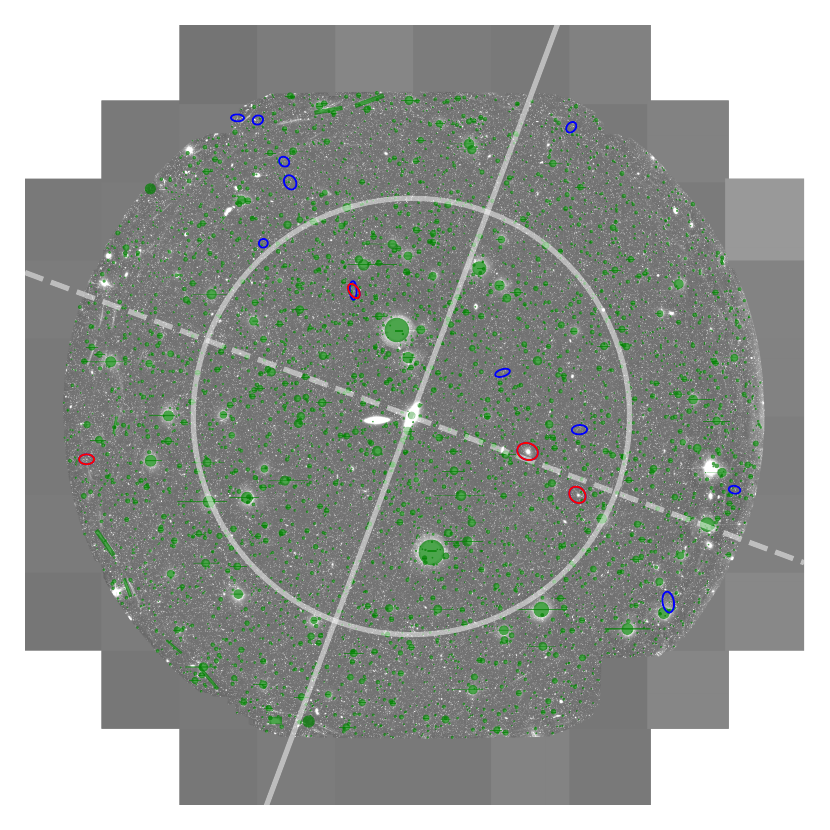}}
    \end{subfigmatrix}
    \caption{
    Spatial distribution of the dwarf galaxies.
    The red and blue circles represent the secure and possible dwarf galaxies, respectively.
    The green shaded circles and boxes show the masked regions.
    The white circle is the virial radius of the host galaxy.
    Note that virial radius of N5866 is outside of the field of view.
    The white solid and dashed lines represent the central galaxy's major and minor axis, respectively.
    }
    \label{fig:images}
\end{figure*}
\twocolumngrid
\subsection{Spatial Distribution}
\label{sec:spa_dist}
Next, we examine the spatial distribution of dwarf satellites shown in Figure \ref{fig:images} for each galaxy.
Since it is difficult to measure the precise distance of dwarf galaxies, we deal with the projected spatial distribution.
We focus on two distributions: radial distribution and angular distribution with respect to the host's major axis.

\subsubsection{Radial Distribution}
\label{sec:rad_dist}
\begin{figure}[t]
    \centering
	\includegraphics[scale=1]{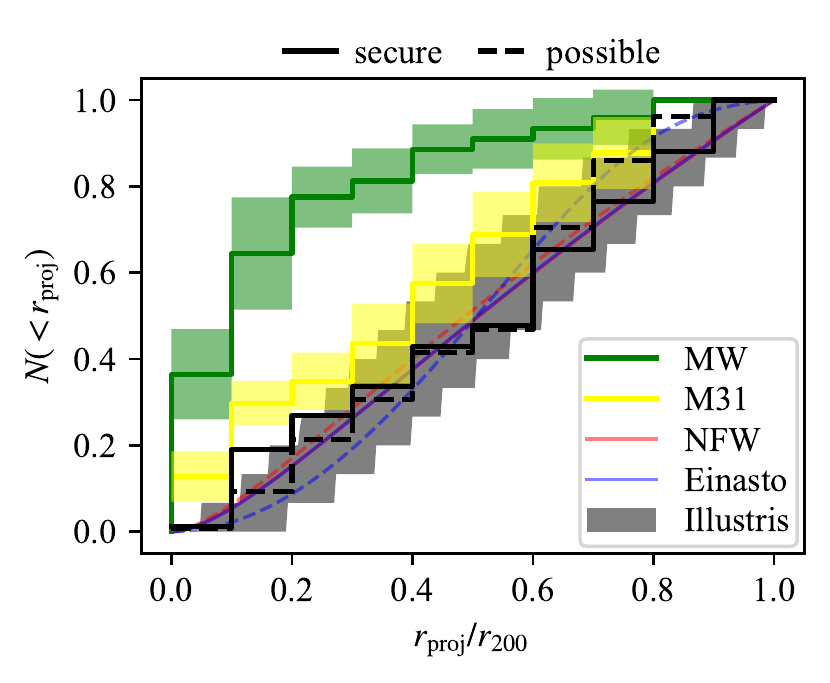}
	\caption{
	    Normalized cumulative distributions of the projected distance from each host galaxy, stacked for all galaxies (black lines), 
	    and the LG satellites around the MW and M31 (green and yellow lines; \citealt{McConnachie12}).
	    For the MW (M31), the green (yellow) shaded region represents the variation of the number count due to projection effects (see text for details).
	    The red and blue lines show the best fit models of the NFW and Einasto profile, respectively.
	    The solid line is for secure satellites and the dashed line includes both secure and possible dwarfs.
	    Note that the blue solid curve overlaps the red solid curve.
	    The shaded region shows the predicted radial distribution with IllustrisTNG-100 \citep{Carlsten+20}.
	    }
    \label{fig:radial}
\end{figure}
Figure \ref{fig:radial} shows the cumulative distribution of the projection distance from the host galaxies to the dwarf satellites normalized by the virial radii and averaged for all the hosts.
We fit the cumulative projected radial distribution of dwarf galaxies with the NFW profile and the Einasto profile \citep{NFW96,Einasto65}, the standard density models, integrated over the line-of-sight direction as in the following equation,
\begin{eqnarray}
    N_\mathrm{NFW}
    &\propto&
    \left\{ \begin{array}{ll}
        \ln \frac{x}{2} 
        + \frac{2}{\sqrt{x^2-1}} \arctan \sqrt{\frac{x-1}{x+1}}
        & (x > 1),
        \\
        1 - \ln 2 
        & (x = 1),
        \\
        \ln \frac{x}{2}
        + \frac{2}{\sqrt{1-x^2}} \mathrm{arctanh} \sqrt{\frac{1-x}{1+x}}
        & (x < 1),
    \end{array} \right. 
    \label{eq:N_nfw} \\
    N_\mathrm{E}
    &\propto&
    \int_0^x ds \, s
    \int^\infty_s dt
    \frac{t}{\sqrt{t^2-s^2}} 
    \exp \left(-\frac{2}{\alpha} t^{\alpha}\right),
    \label{eq:N_einasto} \\
    x &\equiv& c \frac{r_\mathrm{proj}}{r_{200}},
    \label{eq:x}
\end{eqnarray}
where $c$ is the concentration parameter and $\alpha$ is the shape parameter of the Einasto profile.
The parameters of the best fit model is shown in Table \ref{tab:radial_model}. 
Since both models reproduce the data well and have small reduced chi-squares due to the large variance of the sample, it is not easy to judge the relative merits of the two models.
We note that there is a slight difference in the reduced chi-squares of both models in the sample with the possible galaxies.
The trend obtained by both model fitting is that the concentration parameter is between 1 and 2 for any model, which is small compared to the value for the MW ($c \simeq 18-20$: \citealt{Battaglia+05, Catena+10, Deason+12, Nesti+13}).

In order to make a fair comparison between our sample and the MW, we perform mock observations of the projected radial distribution from every projection angle using the 3D positions for MW satellite galaxies brighter than $M_V=-10$ \citep{McConnachie12}.
We find that the radial distribution of MW satellite galaxies shown as the green line is clearly more centrally concentrated.
For the M31, the radial distribution shown as the yellow line is also slightly more concentrated than that of our sample.
In other words, satellite galaxies outside of the LG are more uniformly distributed.
We further compare with the prediction of the numerical simulation shown in \cite{Carlsten+20}, which is based on the cosmological magneto-hydrodynamics simulation, IllustrisTNG-100 \citep{Marinacci+18, Naiman+18, Nelson+18, Pillepich+18, Springel+18}. 
The shaded region, representing the $1\sigma$ spread due to different projection angles, is consistent with our dwarf satellite sample.
\begin{deluxetable}{ccccc}
\tablecaption{Best fit models for projected radial profiles.}
\tablehead{
& \multicolumn{2}{c}{secure} & \multicolumn{2}{c}{with possible} \\
& \colhead{NFW} & \colhead{Einasto} & \colhead{NFW} & \colhead{Einasto}}
\startdata
$c$      & 1.67$\pm$0.32 & 1.60$\pm$0.36 & 2.11$\pm$0.72 & 1.43$\pm$0.12 \\ 
$\alpha$ & ---           & 0.26$\pm$0.16 & --- & 5.93$\pm$3.81 \\
$\chi^2$\tablenotemark{a} & 0.028 & 0.031 & 0.285 & 0.090
\enddata
\tablenotetext{a}{Reduced chi-square.}
\label{tab:radial_model}
\end{deluxetable}

\cite{Mao+21} reported that the average radial distribution of galaxies at distances greater than 25 Mpc obtained from the SAGA survey is not centrally concentrated and agrees with the simulation predictions.
Their result is consistent with our finding here; the distribution of the satellite galaxies around the nine MW-like host galaxies is less concentrated than that of the MW satellites.
Among our secure satellite samples (we exclude N5470 from statistical analyses because N5470 has no secure satellite), the median and the 16th/84th percentiles of the median projected distance within 150 kpc of the projected distance is $d_\mathrm{proj,half}^{(<150\,\mathrm{kpc})}=104^{+20}_{-59}$ kpc, which is consistent with the result of the SAGA survey, $d_\mathrm{proj,half}^{(<150\,\mathrm{kpc})}=83^{+48}_{-36}$ kpc \citep{Mao+21}. 
This result also agrees with the prediction by the simulation in \cite{Carlsten+20} within 1$\sigma$. 
Taking into account the possible satellite samples does not affect these results ($d_\mathrm{proj,half}^{(<150\mathrm{kpc})}=101^{+28}_{-47}$ kpc). 
For the MW, it is estimated to $d_\mathrm{proj,half}^{(<150\,\mathrm{kpc})}=43^{+11}_{-16}$ kpc from \cite{McConnachie12}, which is less than our and SAGA results. 
The characteristic radial distribution of the MW dwarf galaxies might have something to do with the existence of the Large and Small Magellanic Clouds and the unique plane structure known as the vast polar structure (VPOS; see Section \ref{sec:ang_dist}).

\subsubsection{Angular Distribution}
\label{sec:ang_dist}
\begin{figure}[t]
    \centering
	\includegraphics[scale=1]{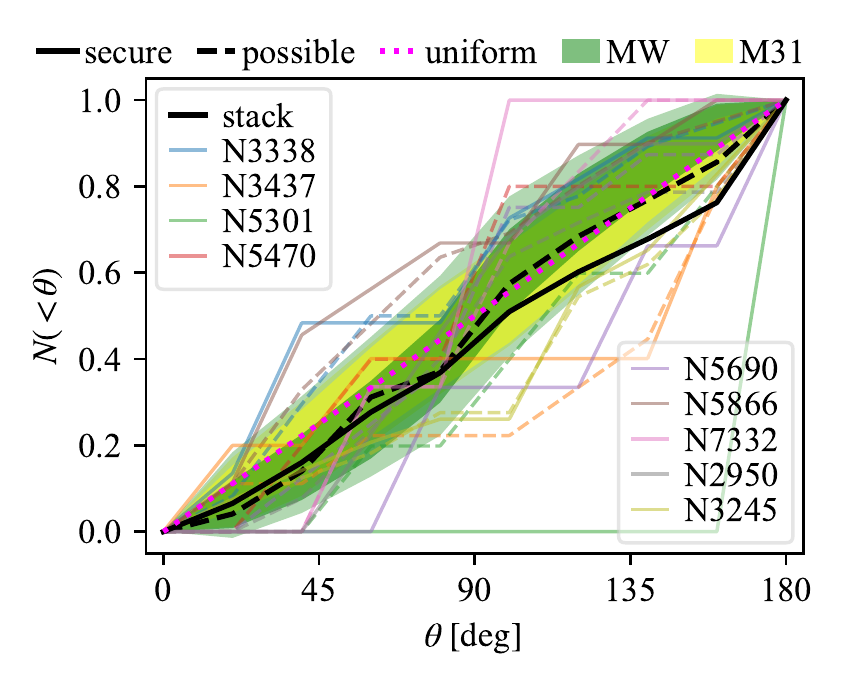}
	\caption{
	    Normalized cumulative angular distribution of dwarf galaxies.
	    Solid and dashed lines represent observed dwarf galaxies divided into the secure and possible group, respectively. 
	    The black line is for stacked all galaxies, and the colored lines are for each galaxy.
	    The dark and light green (yellow) shaded regions indicate the variation of the MW (M31) satellite galaxies and only those brighter than $M_V=-10$ \citep{McConnachie12}, respectively, which are taken into account the projection effects (see text for details). 
	    The magenta dotted line is estimated using random points in the 3D uniform distribution.
	    }
    \label{fig:angular}
\end{figure}
We move on to examine the distribution as a function of the angular separation.
Dwarf galaxies in the LG have an enhanced distribution in the plane relative to the host galaxy, such as the VPOA in the MW \citep{Pawlowski+12} and the great plane of Andromeda (GPoA) in the Andromeda galaxy \citep{Ibata+13, Conn+13}.
If our sample of dwarf galaxies has such a peculiar planar structure, it is expected to show an angular distribution enhanced to a specific angle.

Figure \ref{fig:angular} shows the position angle distribution of the dwarf galaxies measured from the major axis of each host galaxy.
The deviations of individual lines from the uniform distribution shown as the dotted line can be interpreted as a sign of alignment.
We first focus on individual galaxies.  It is evident that there is a large host-host scatter; some galaxies are close to the uniform
distribution while others show significant spatial inhomogeneity in the satellite distribution.  However, we find it difficult to
interpret the deviations observed here for two reasons.  One is because we apply large masks around bright stars, which affect the angular distribution of the satellites.
We account for the mask as described in Section 3.5, but the mask does increase the scatter.
The other is that some of the host galaxies have only a small number of satellites (e.g., N7732) and the scatter is likely due to a statistical fluctuation.

To mitigate these two effects, we stack all the host galaxies as shown by the thick lines in Figure \ref{fig:angular}.
We find that the angular distribution is consistent with a uniform distribution.
For comparison, we also show the satellite distributions of the MW and M31 as the shaded area.
We estimate the angular distributions of MW and M31 by mock observations based on the 3D position data in \cite{McConnachie12} as in Section \ref{sec:rad_dist}. 
In the mock observation, we determine the major axis of the host galaxies viewed from a given angle assuming that the host galaxies' disk is a perfect circle (for the M31, we use the observed axial ratio and position angle given by \cite{Jarrett+03} to define the direction of the galaxy disk). 
However, the variation of the angular distribution is estimated except in the case of face-on view since the major axis cannot be determined when the inclination angle is zero.
As can be seen from this figure, the MW satellite distribution is S-shaped, which corresponds to the observed alignment.
Because we project the satellite distribution onto 2D, the alignment is less clear depending on the viewing angle and
that results in the somewhat weak trend in the figure.
This signature also appears weakly only in the MW satellites with $M_V < -10$.
For M31, on the other hand, the angular distribution shows no significant alignment like MW, and we cannot distinguish it from a uniform distribution.
Comparisons between our measurements and MW/M31 suggest that the satellite alignment around the external MW-like hosts is not as strong as that observed in MW.

However, we note that it is possible that the alignment direction is independent of the host's major axis and the stacking analysis here may be
averaging out alignments around individual host.  Unfortunately, individual host is difficult to interpret for the reasons described above.
If we could identify much fainter satellites and increase the statistics, we might be able to address the alignment around individual host.

The angular distribution is another possible difference from the MW\@.  
Despite the similarity in luminosity function, the spatial distribution of the MW satellite may be peculiar.  
Before we discuss this point further, we investigate other physical properties of the satellites such as the size-luminosity and color-magnitude relations in the following subsections to see whether there are other differences.

\subsection{Size-Luminosity Relation}
\label{sec:size-lumi}
\begin{figure}[t]
    \centering
	\includegraphics[scale=1]{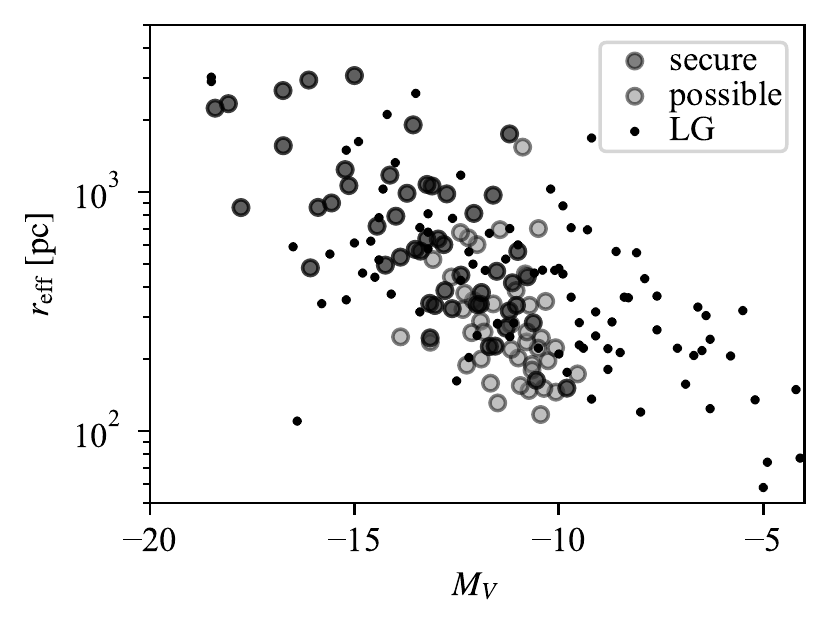}
	\caption{
	    Effective radius versus absolute magnitude.
	    The dark and light gray circles are the secure and possible satellites, respectively. 
	    The dots indicate the satellite galaxies in the LG provided by \cite{McConnachie12}.
	    }
    \label{fig:size_lumi}
\end{figure}
Figure \ref{fig:size_lumi} shows the size-luminosity relation of the dwarf satellites.
For comparison, the satellite galaxies in the LG are overlaid \citep{McConnachie12}.
As can be seen from this figure, our sample shows a similar trend to the LG satellites.
There is one bright and compact satellite in the LG, as plotted in the lower left of the figure.
It is M32 and its compact size is likely due to a tidal disruption.
We miss such a compact source, but it is a fairly minor population of the entire satellites and we do not expect it does not significantly alter our results here.
Dwarf satellite galaxies exhibit a clear correlation between size and magnitude in
the sense that brighter galaxies are bigger.  
Galaxies both in and out of the LG follow the same relation.  
This suggests that the size-luminosity relation is universal, regardless of whether inside or outside the LG\@.

\subsection{Color-Magnitude Relation}
\label{sec:CM_relation}
\begin{figure}[t]
    \centering
	\includegraphics[scale=1]{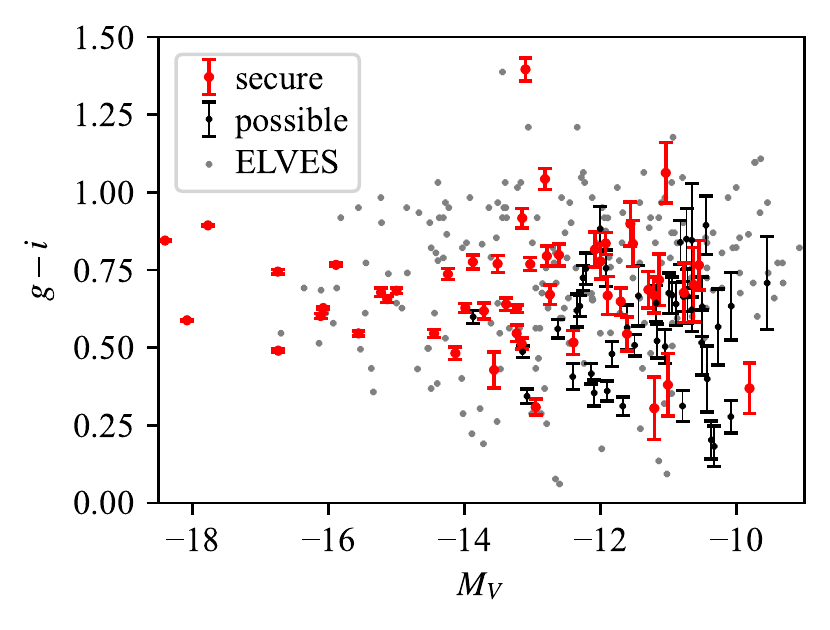}
	\caption{
	    $g-i$ color plotted against absolute magnitude.
	    The red and black circles are the secure and possible satellites, respectively. 
	    The gray dots show the satellite galaxies outside the LG provided by the ELVES survey \citep{Carlsten+21}.
	    }
    \label{fig:CMD}
\end{figure}
We present in Figure \ref{fig:CMD} the color-magnitude diagram of the satellite galaxies.
Most of the bright satellites have a relatively red color with $g-i \sim 0.7$, while faint satellites ($M_V \gtrsim -13$) span a wider color range.
The possible satellites show a relatively higher fraction of the blue galaxies than the secure dwarf galaxies, which might be due to contaminant galaxies.
There is a very red galaxy with $M_V \sim -13$.  
This red color is likely due to contamination of very nearby bright star.
As can be seen from this figure, the overall galaxy distribution is similar to that observed in \cite{Carlsten+21}.
They used CFHT to measure the $g-i$ color, and we translate it into the HSC $g-i$ color using $(g-i)_\mathrm{HSC}=1.057(g-i)_\mathrm{CFHT}-2.232\times10^{-3}$, which is estimated in the same manner as Equation \eqref{eq:Mv}.
The color distribution of the satellites from \cite{Carlsten+21} is overall similar to ours.

\section{Summary and Conclusion} \label{sec:summary}
In this paper, we have performed statistical analyses of satellite galaxies around nine nearby galaxies with MW-like mass observed with HSC on Subaru.
Specifically, we have examined the luminosity function, spatial distribution, size-mass relation and color-magnitude diagram and compared them with the MW satellites and satellites of other galaxies outside the LG from the literature. 

Our analyses are based on the careful selection of the satellite candidates: 
\begin{inparaenum}[(1)]
    \item 
    multi-path selection of satellites using various observables such as surface brightness, magnitude, and S\'{e}rsic index using Source Extractor and GALFIT, 
    \item 
    SBF distance measurements to further reduce contamination, and 
    \item 
    finally careful visual inspection to remove remaining artifacts and classify candidates into secure and possible satellites.
\end{inparaenum}
Following this procedure, we have identified 51 (42) secure (possible) satellite galaxies within the virial radius of nine nearby galaxies, including the results of pilot observations \citepalias{Tanaka+18}.
The main results based on the satellite galaxy catalog can be summarized as follows.
\begin{enumerate}
   \setlength{\itemsep}{-1mm}
   \setlength{\leftskip}{-1mm}
   \item 
   The average luminosity function is very similar to that of the MW satellites.
   Compared to the luminosity function of more distant satellites from \cite{Wang+21}, there is a slight deviation at faint magnitudes ($M_V\sim-10$), but it is within the scatter.
   The large error bars in the luminosity function indicates that the distribution of the dwarf galaxy population is highly variable from galaxy to galaxy.  
   This scatter can be caused in part by the host mass dependence as our sample shows a strong correlation between the number of satellites and the host halo mass.
   But, there also seems to be a host-host scatter at fixed mass.
   \item 
   We show that the size-luminosity relation of our satellites is consistent with that of the LG, suggesting that the relation is universal.
   \item 
   In contrast to these photometric properties, there is a clear difference
   in the spatial distribution of satellites.
   We find no sign of angular alignments in the projected distribution of the satellites, and the satellites are distributed rather uniformly inside the virial radius.  
   On the other hand, the LG satellites are aligned (VPOA and SPoA) and are more centrally concentrated around the hosts.
\end{enumerate}
These results suggest similar photometric properties but different spatial distributions of the satellite galaxies in and out of the LG.

The observed difference in the radial distribution of satellites is unlikely due to observational incompleteness; our observation covers the entire virial radius of the hosts (except for N5866) and is highly complete.  
The only region we are missing satellites is the central $\sim10 kpc$ region, where the host galaxy dominates and it is hard to identify diffuse satellites.  
We do not expect that the incompleteness in such a small region affects our conclusion in a significant way.  
As for the MW satellites, previous observations are deep enough to discover all massive satellites with $M_V<-10$ as discussed in \cite{Carlsten+20}.  
Therefore, the observed difference in the radial distribution is likely real.

The missing satellite problem is about the satellite abundance, but as we have shown in this paper, the MW satellite abundance is within the scatter of the abundance observed in other galaxies.  
Recent simulations also seem to roughly reproduce the observed abundance of the MW satellites.
However, there now seems to be a problem in the spatial distribution in the sense that the MW satellite are too centrally concentrated.
The MW may not be a typical galaxy in this respect, and in fact, there are indications that the MW is not typical (e.g., \citealt{mutch11,tollerud11,rodriguez13}).
While the MW is the closest galaxy and we can study it in great detail, care should be taken when using the MW and its satellites to test galaxy formation models and cosmological problems.
At this point, we can only make speculations about the origin of the centrally concentrated satellite distribution (e.g., peculiarity in the accretion history), and further work is needed to understand its physical origin.

In this paper, we have conservatively classified all dwarf-like objects as possible satellite galaxies not to miss real satellites by the visual inspection.
Our sample includes similar numbers of possible satellites to the secure satellites.
Most of the possible satellites are faint galaxies, and not all the possible satellites are real satellites and there remains a level of contamination of fore/background galaxies.
It is hard to further screen the satellite candidates with photometric data alone.
While we do not expect that the contamination will change our main conclusion (i.e., radial distribution), follow-up spectroscopic observations will be essential.
The Prime Focus Spectrograph \citep{Tamura+16} to be deployed on the Subaru telescope is the best instrument for such follow-up observations because PFS covers a similar field to that of the HSC\@.
We hope to perform such follow-up observations to securely identify the satellites in our future work.

\section*{Acknowledgements}
This research is based on data collected at the Subaru Telescope, which is operated by the National Astronomical Observatory of Japan. We are honored and grateful for the opportunity of observing the Universe from Maunakea, which has the cultural, historical, and natural significance in Hawaii. 

This work is supported by JSPS KAKENHI Grant Numbers 19H01931, 19H01942, 20H01895, 20H05861, 21K13909, 21H04496, 21H05447, and 21H05448. 
 
 The Hyper Suprime-Cam (HSC) collaboration includes the astronomical communities of Japan and Taiwan, and Princeton University. The HSC instrumentation and software were developed by the National Astronomical Observatory of Japan (NAOJ), the Kavli Institute for the Physics and Mathematics of the Universe (Kavli IPMU), the University of Tokyo, the High Energy Accelerator Research Organization (KEK), the Academia Sinica Institute for Astronomy and Astrophysics in Taiwan (ASIAA), and Princeton University. Funding was contributed by the FIRST program from the Japanese Cabinet Office, the Ministry of Education, Culture, Sports, Science and Technology (MEXT), the Japan Society for the Promotion of Science (JSPS), Japan Science and Technology Agency (JST), the Toray Science Foundation, NAOJ, Kavli IPMU, KEK, ASIAA, and Princeton University. 

 This paper makes use of software developed for Vera C. Rubin Observatory. We thank the Rubin Observatory for making their code available as free software at \url{http://pipelines.lsst.io/}.

 This paper is based on data collected at the Subaru Telescope and retrieved from the HSC data archive system, which is operated by the Subaru Telescope and Astronomy Data Center (ADC) at NAOJ\@. Data analysis was in part carried out with the cooperation of Center for Computational Astrophysics (CfCA), NAOJ\@.

 Funding for the SDSS and SDSS-II has been provided by the Alfred P. Sloan Foundation, the Participating Institutions, the National Science Foundation, the U.S. Department of Energy, the National Aeronautics and Space Administration, the Japanese Monbukagakusho, the Max Planck Society, and the Higher Education Funding Council for England. The SDSS Web Site is \url{http://www.sdss.org/}.

 The SDSS is managed by the Astrophysical Research Consortium for the Participating Institutions. The Participating Institutions are the American Museum of Natural History, Astrophysical Institute Potsdam, University of Basel, University of Cambridge, Case Western Reserve University, University of Chicago, Drexel University, Fermilab, the Institute for Advanced Study, the Japan Participation Group, Johns Hopkins University, the Joint Institute for Nuclear Astrophysics, the Kavli Institute for Particle Astrophysics and Cosmology, the Korean Scientist Group, the Chinese Academy of Sciences (LAMOST), Los Alamos National Laboratory, the Max-Planck-Institute for Astronomy (MPIA), the Max-Planck-Institute for Astrophysics (MPA), New Mexico State University, Ohio State University, University of Pittsburgh, University of Portsmouth, Princeton University, the United States Naval Observatory, and the University of Washington.

\software{GALFIT \citep{galfit02, galfit10}, 
          Source Extractor \citep{SExtractor96}
          }

\appendix
\section{Satellite Galaxy Catalog} \label{sec:catalog}
\begin{figure*}[t]
    \centering
    \includegraphics[width=0.8\linewidth]{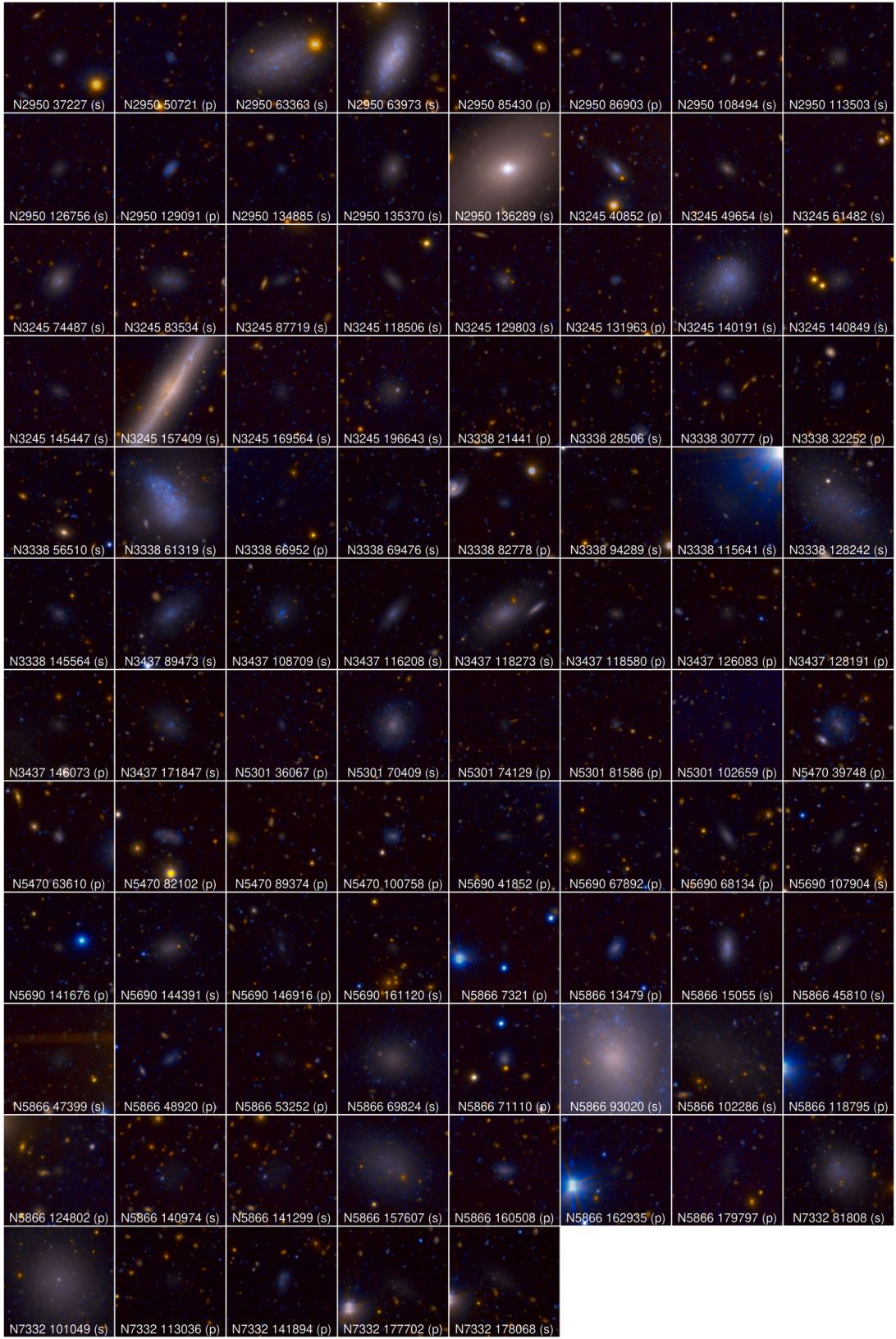}
    \caption{Images of 93 satellite galaxies. The label on each image shows the host galaxy, the serial number of detected objects, and the confidence flag, which are the same as Table \ref{tab:satellite_list}. The each cutout size is 1 arcmin $\times$ 1 arcmin.
    }
    \label{fig:montage}
\end{figure*}

Here we present satellite galaxy catalog (Table \ref{tab:satellite_list}) and images (Figure \ref{fig:montage}).

\startlongtable
\begin{deluxetable*}{ccr@{.}lr@{.}lcccccc}
\tablecaption{The list of satellite galaxies.}
\tablehead{
\colhead{Host} & \colhead{Object ID} & \multicolumn{2}{c}{R.A.} & \multicolumn{2}{c}{Dec} & \colhead{$m_i$} & \colhead{$\mu_i$} & \colhead{$g-i$} &\colhead{$M_V$} & \colhead{$n$} & \colhead{confidence} \\
& & \multicolumn{2}{c}{(deg)} & \multicolumn{2}{c}{(deg)} & \colhead{(mag)} & \colhead{($\mathrm{mag\, arcsec^{-2}}$)} & & \colhead{(mag)} & & } 
\startdata
N2950 & 37227 & 145&3691 & +58&4171 & 19.59 & 23.64 & 0.65 & -11.70 & 1.02 & s \\
N2950 & 50721 & 146&1931 & +58&5106 & 20.66 & 23.99 & 0.20 & -10.37 & 0.50 & p \\
N2950 & 63363 & 145&0083 & +58&5353 & 15.47 & 22.71 & 0.77 & -15.89 & 0.67 & s \\
N2950 & 63973 & 145&8459 & +58&5245 & 15.20 & 21.42 & 0.63 & -16.08 & 0.80 & s \\
N2950 & 85430 & 146&7232 & +58&7216 & 18.05 & 22.77 & 0.49 & -13.14 & 0.92 & p \\
N2950 & 86903 & 146&7853 & +58&7427 & 20.36 & 23.72 & 0.67 & -10.94 & 0.89 & p \\
N2950 & 108494 & 144&7900 & +58&8852 & 20.81 & 24.18 & 0.77 & -10.55 & 0.84 & s \\
N2950 & 113503 & 145&6677 & +58&9129 & 19.48 & 24.17 & 0.84 & -11.92 & 0.73 & s \\
N2950 & 126756 & 146&4625 & +58&9936 & 20.02 & 24.65 & 0.69 & -11.29 & 0.95 & s \\
N2950 & 129091 & 145&7567 & +59&0110 & 19.42 & 23.19 & 0.31 & -11.67 & 0.51 & p \\
N2950 & 134885 & 145&6686 & +59&0501 & 21.32 & 24.39 & 0.37 & -9.80 & 0.66 & s \\
N2950 & 135370 & 145&5150 & +59&0429 & 18.33 & 23.50 & 0.77 & -13.03 & 1.12 & s \\
N2950 & 136289 & 145&7577 & +58&9736 & 13.66 & 21.50 & 0.89 & -17.78 & 3.06 & s \\
\hline
N3245 & 40852 & 156&8810 & +28&1036 & 18.12 & 22.57 & 0.60 & -13.87 & 1.20 & p \\
N3245 & 49654 & 157&0612 & +28&1659 & 19.03 & 22.86 & 0.92 & -13.15 & 1.41 & s \\
N3245 & 61482 & 156&4155 & +28&2391 & 20.61 & 24.05 & 0.90 & -11.56 & 0.97 & s \\
N3245 & 74487 & 156&7343 & +28&3066 & 17.83 & 22.97 & 0.74 & -14.24 & 0.80 & s \\
N3245 & 83534 & 157&2669 & +28&3654 & 18.22 & 24.01 & 0.78 & -13.88 & 0.75 & s \\
N3245 & 87719 & 157&1272 & +28&4015 & 20.10 & 24.73 & 0.78 & -12.00 & 0.94 & s \\
N3245 & 118506 & 157&1483 & +28&5841 & 19.50 & 23.97 & 0.80 & -12.61 & 0.60 & s \\
N3245 & 129803 & 156&4046 & +28&6472 & 19.32 & 23.86 & 0.79 & -12.79 & 1.24 & s \\
N3245 & 131963 & 157&2335 & +28&6723 & 20.78 & 23.94 & 0.52 & -11.17 & 0.50 & p \\
N3245 & 140191 & 156&4141 & +28&6801 & 16.40 & 22.80 & 0.55 & -15.56 & 0.73 & s \\
N3245 & 140849 & 156&3867 & +28&7026 & 19.44 & 24.34 & 1.04 & -12.81 & 1.35 & s \\
N3245 & 145447 & 156&8121 & +28&7554 & 20.14 & 24.42 & 0.67 & -11.89 & 0.82 & s \\
N3245 & 169564 & 156&6434 & +29&0628 & 20.61 & 25.59 & 0.83 & -11.52 & 1.01 & s \\
N3245 & 196643 & 156&6751 & +28&8594 & 18.58 & 24.70 & 0.77 & -13.51 & 1.11 & s \\
N3245 & 157409 & 156&7550 & +28&6392 & 13.73 & 21.13 & 0.84 & -18.41 & 1.18 & s \\
\hline
N3338 & 21441 & 160&8573 & +13&2273 & 21.95 & 24.64 & 0.57 & -10.27 & 0.92 & p \\
N3338 & 28506 & 160&4188 & +13&2796 & 21.51 & 25.90 & 0.68 & -10.77 & 0.91 & s \\
N3338 & 30777 & 160&0761 & +13&2946 & 19.90 & 24.10 & 0.62 & -12.35 & 0.97 & p \\
N3338 & 32252 & 160&4134 & +13&3044 & 20.00 & 24.11 & 0.35 & -12.09 & 0.50 & p \\
N3338 & 56510 & 160&3419 & +13&4713 & 21.67 & 25.55 & 0.70 & -10.63 & 0.79 & s \\
N3338 & 61319 & 160&7730 & +13&5117 & 15.43 & 22.70 & 0.49 & -16.74 & 0.79 & s \\
N3338 & 66952 & 160&1169 & +13&5356 & 21.69 & 25.07 & 0.40 & -10.43 & 1.02 & p \\
N3338 & 69476 & 160&3212 & +13&5517 & 21.10 & 26.31 & 0.38 & -11.01 & 1.14 & s \\
N3338 & 82778 & 160&6315 & +13&6357 & 20.61 & 24.79 & 0.56 & -11.60 & 1.52 & p \\
N3338 & 94289 & 160&3061 & +13&7023 & 21.17 & 25.57 & 0.72 & -11.14 & 0.69 & s \\
N3338 & 115641 & 160&5417 & +13&8261 & 18.57 & 26.21 & 0.43 & -13.56 & 1.10 & s \\
N3338 & 128242 & 159&9818 & +13&9093 & 16.12 & 24.56 & 0.60 & -16.12 & 1.90 & s \\
N3338 & 145564 & 160&1787 & +13&9995 & 19.79 & 24.45 & 0.52 & -12.40 & 1.29 & s \\
\hline
N3437 & 89473 & 163&5448 & +22&8294 & 17.99 & 24.44 & 0.48 & -14.13 & 1.18 & s \\
N3437 & 108709 & 163&0100 & +22&9405 & 19.07 & 24.58 & 0.31 & -12.95 & 1.11 & s \\
N3437 & 116208 & 163&6784 & +22&9738 & 18.93 & 23.87 & 0.55 & -13.23 & 1.08 & s \\
N3437 & 118273 & 163&1291 & +23&0015 & 17.09 & 23.34 & 0.66 & -15.14 & 1.02 & s \\
N3437 & 118580 & 163&6385 & +22&9955 & 20.37 & 24.03 & 0.76 & -11.92 & 0.97 & p \\
N3437 & 126083 & 163&6439 & +23&0328 & 20.36 & 25.82 & 0.88 & -12.00 & 1.37 & p \\
N3437 & 128191 & 162&6322 & +23&0457 & 21.61 & 24.10 & 0.85 & -10.73 & 0.90 & p \\
N3437 & 146073 & 163&4759 & +23&1417 & 20.07 & 25.53 & 0.75 & -12.21 & 1.64 & p \\
N3437 & 171847 & 162&6968 & +23&2901 & 18.49 & 24.83 & 0.62 & -13.71 & 1.23 & s \\
\hline
N5301 & 36067 & 206&5198 & +45&7738 & 21.40 & 24.60 & 0.52 & -10.51 & 0.60 & p \\
N5301 & 70409 & 206&6770 & +46&0531 & 17.48 & 23.54 & 0.55 & -14.45 & 1.41 & s \\
N5301 & 74129 & 206&6924 & +46&0824 & 21.28 & 25.49 & 0.67 & -10.72 & 0.70 & p \\
N5301 & 81586 & 206&7805 & +46&1407 & 21.01 & 24.31 & 0.68 & -10.99 & 0.73 & p \\
N5301 & 102659 & 206&9393 & +46&3099 & 21.45 & 24.82 & 0.85 & -10.65 & 0.52 & p \\
\hline
N5470 & 39748 & 211&8306 & +5&7113 & 18.70 & 24.10 & 0.34 & -13.08 & 1.43 & p \\
N5470 & 63610 & 211&4875 & +5&8939 & 19.75 & 22.77 & 0.72 & -12.25 & 0.69 & p \\
N5470 & 82102 & 211&5376 & +6&0315 & 19.28 & 23.64 & 0.56 & -12.63 & 0.67 & p \\
N5470 & 89374 & 211&4673 & +6&0844 & 21.66 & 23.92 & 0.89 & -10.45 & 0.53 & p \\
N5470 & 100758 & 211&5476 & +6&1575 & 19.69 & 23.32 & 0.42 & -12.13 & 0.71 & p \\
\hline
N5690 & 41852 & 219&3252 & +1&9058 & 20.64 & 24.35 & 0.64 & -11.18 & 1.05 & p \\
N5690 & 67892 & 219&0349 & +2&0722 & 21.74 & 24.60 & 0.63 & -10.07 & 0.82 & p \\
N5690 & 68134 & 219&1551 & +2&0663 & 19.51 & 23.59 & 0.63 & -12.30 & 0.93 & p \\
N5690 & 107904 & 219&6810 & +2&3119 & 20.62 & 24.81 & 0.67 & -11.21 & 0.86 & s \\
N5690 & 141676 & 219&3985 & +2&5156 & 21.15 & 24.39 & 0.62 & -10.66 & 1.41 & p \\
N5690 & 144391 & 219&0838 & +2&5217 & 18.43 & 23.96 & 0.64 & -13.39 & 0.80 & s \\
N5690 & 146916 & 219&7186 & +2&5426 & 20.69 & 25.08 & 0.50 & -11.05 & 1.81 & p \\
N5690 & 161120 & 219&6768 & +2&6338 & 21.03 & 25.54 & 1.06 & -11.04 & 1.08 & s \\
\hline
N5866 & 7321 & 227&0669 & +55&1291 & 21.76 & 25.62 & 0.71 & -9.54 & 0.91 & p \\
N5866 & 13479 & 226&3043 & +55&2006 & 19.20 & 22.90 & 0.36 & -11.90 & 0.98 & p \\
N5866 & 15055 & 226&7565 & +55&2240 & 18.03 & 22.60 & 0.51 & -13.16 & 0.73 & s \\
N5866 & 45810 & 226&8188 & +55&4747 & 18.54 & 25.33 & 0.67 & -12.74 & 3.04 & s \\
N5866 & 47399 & 225&7834 & +55&4831 & 18.61 & 26.27 & 1.40 & -13.10 & 1.43 & s \\
N5866 & 48920 & 225&6471 & +55&4993 & 20.28 & 24.19 & 0.31 & -10.79 & 1.10 & p \\
N5866 & 53252 & 226&4683 & +55&5333 & 20.56 & 26.55 & 0.84 & -10.82 & 1.83 & p \\
N5866 & 69824 & 226&2083 & +55&6451 & 17.27 & 24.46 & 0.63 & -13.99 & 1.56 & s \\
N5866 & 71110 & 225&7158 & +55&6481 & 19.69 & 22.90 & 0.51 & -11.50 & 0.57 & p \\
N5866 & 93020 & 228&0289 & +55&7848 & 13.15 & 22.38 & 0.59 & -18.08 & 1.59 & s \\
N5866 & 102286 & 226&3762 & +55&8661 & 16.29 & 25.53 & 0.68 & -15.00 & 1.00 & s \\
N5866 & 118795 & 226&5328 & +55&9595 & 20.97 & 24.98 & 0.28 & -10.08 & 0.83 & p \\
N5866 & 124802 & 225&9643 & +56&0019 & 20.50 & 25.02 & 0.66 & -10.78 & 1.39 & p \\
N5866 & 140974 & 226&3328 & +56&1180 & 19.86 & 28.44 & 0.30 & -11.21 & 1.71 & s \\
N5866 & 141299 & 226&3304 & +56&1178 & 19.60 & 27.05 & 0.54 & -11.61 & 2.71 & s \\
N5866 & 157607 & 227&0206 & +56&2583 & 18.03 & 25.26 & 0.63 & -13.23 & 1.33 & s \\
N5866 & 160508 & 225&9496 & +56&2563 & 19.34 & 23.65 & 0.48 & -11.83 & 1.02 & p \\
N5866 & 162935 & 227&3021 & +56&2795 & 20.67 & 25.91 & 0.18 & -10.32 & 1.58 & p \\
N5866 & 179797 & 226&5513 & +56&4816 & 19.84 & 26.40 & 0.67 & -11.44 & 1.75 & p \\
\hline
N7332 & 81808 & 338&9189 & +23&6076 & 17.02 & 24.04 & 0.68 & -15.23 & 0.88 & s \\
N7332 & 101049 & 339&0489 & +23&7117 & 15.54 & 24.16 & 0.74 & -16.75 & 1.84 & s \\
N7332 & 113036 & 338&9130 & +23&7637 & 21.72 & 27.16 & 0.63 & -10.50 & 1.39 & p \\
N7332 & 141894 & 339&1142 & +23&9002 & 19.68 & 24.77 & 0.41 & -12.41 & 2.05 & p \\
N7332 & 177702 & 339&5058 & +24&0978 & 21.34 & 27.96 & 0.64 & -10.88 & 2.91 & p \\
N7332 & 178068 & 339&5038 & +24&0965 & 20.25 & 25.97 & 0.82 & -12.08 & 0.81 & s \\
\enddata
\tablecomments{The columns list the host galaxies, serial numbers of detected objects, coordinates, apparent magnitudes in $i$-band, surface brightness in $i$-band, colors, absolute magnitudes in $V$-band estimated by Equation \eqref{eq:Mv}, S\'{e}rsic indices, and confidence flags where s and p stand for secure and possible satellite galaxy, respectively. Colors are given by the Source Extractor and other observables by the Galfit measurements.}
\label{tab:satellite_list}
\end{deluxetable*}

\end{document}